\documentclass[aps]{revtex4}

\usepackage{graphicx}
\usepackage{epsfig}
\usepackage{dcolumn}
\usepackage{bm}

\begin{document}

\preprint{}

\title{Nonlinear Kinetic Development of the Weibel Instability \\ and the generation of electrostatic coherent structures}
\author{L. Palodhi}
\email{lopamudra@df.unipi.it} \affiliation{Physics Department, University of Pisa, Pisa, Italy }
\author{F. Califano}
\email{califano@df.unipi.it} 
\author{F. Pegoraro}
\email{pegoraro@df.unipi.it} \affiliation{Physics Department and CNISM, University of Pisa, Pisa, Italy }

\begin{abstract}

The nonlinear evolution of the Weibel instability  driven   by the anisotropy of the electron distribution function  in a
collisionless plasma is investigated in a spatially one-dimensional configuration with a Vlasov code in a two-dimensional
velocity space.  It is found that the electromagnetic fields generated by this  instability cause  a  strong deformation of the
electron distribution function in phase space, corresponding to highly filamented  magnetic vortices. Eventually, these deformations  lead to the generation of short wavelength Langmuir modes  that form highly  localized electrostatic structures corresponding to jumps of the electrostatic potential.

\end{abstract}


\maketitle

\section{Introduction}
\label{introduction}

The generation of macroscopic magnetic fields in a plasma is a fundamental process that occurs in  laboratory and astrophysical
plasma  configurations in  a wide range of regimes of plasma collisionality. This process appears naturally in a plasma. In
plasma dynamics the  ``mechanical'' and  ``electrodynamic''  degrees of freedom couple very  efficiently allowing for a rapid
transfer  from mechanical to electrodynamic energy and vice versa, as exemplified by the the dynamo mechanism \cite{DYNREV} which is responsible for the generation of the magnetic fields of the planets and of  the stars.

Here we are interested in investigating the different physical mechanisms that can accompany  the generation of magnetic fields in a collisionless  plasma regime,  i.e., under  conditions  that are very far from thermodynamic equilibrium,  and in particular in the generation of  longitudinal electric fields and  of electrostatic potential jumps due to nonlinear effects and  to  the related modification of the   electron distribution function in phase space that accompany the generation of the magnetic field.

We refer in particular to  the basic mechanism identified more than half a century ago by Weibel \cite{WEIBEL} where the magnetic field is generated by  the development of an instability driven  by the temperature anisotropy of  the electron distribution function  in a  collisionless plasma. Later on, a systematic linear kinetic analysis of the instabilities at play in an anisotropic plasma, without an initial mean magnetic field, has been carried out in an electron-ion plasma \cite{kalman}. We recall that other  mechanisms of magnetic field generation can be at play in different collisionality regimes, as extensively reviewed  in Ref.  \cite{HAINES}.

A rough  physical   model  of the magnetic field generation can be formulated in terms of a ``thermal'' machine working between the two 'temperatures', $T_x$ and $T_y$ which correspond to the two 'temperatures' that characterize the electron distribution function:  $T_y$ describes the distribution  width in  $v_y$ and $T_x < T_y$ the distribution  width in $v_x$.
In a highly collisional regime the two temperatures would be rapidly equalized by  the effect of collisions. In a collisionless plasma regime work can be extracted from this 'temperature' difference. That this work is associated with the generation of magnetic field, i.e. in other words that this work corresponds to the transformation of thermal energy into magnetic energy, can be guessed by means of a virtual displacement argument borrowed from the theory of the closely related current filamentation instability\cite{CFI}.
We imagine splitting at each point the electron distribution function into two parts,
corresponding to positive and to negative values of $v_y$ respectively, and displacing them in the transverse plane  with  a virtual displacement of the form    $ \xi^+_{0x} \sin{(k_x x)}  =  - \xi^-_{0x} \sin{(k_x x)}$, where the upper indices refer to the  populations with positive and with negative value of $v_y$, respectively. In this way opposite current densities along $y$ are formed  in the plasma. These currents  are modulated along $x$ and produce a magnetic field along $z$.  Since  opposite currents repel, the initial displacement is reinforced,  the instability can develop and the magnetic field along $z$ can grow.

Actually, the nonlinear evolution of the Weibel instability  is far richer, even when we focus our investigation  on a simplified geometrical configuration  where all quantities are assumed to depend only on $x$ and velocity space is reduced  to the $v_x$-$v_y$  plane.  In this context we recall that the restriction to such a 1D-2V configuration  excludes, among other phenomena that can accompany the nonlinear evolution  of the Weibel instability,  the development of a secondary magnetic field line reconnection instability  of the type discussed in Ref.   \cite{CALIFANOR}.

In the one dimensional configuration considered here,  the development of the Weibel instability   is accompanied by the generation of electric fields: an inductive electric field in the $y$ direction intrinsically related to the variation in time of $B_z$ and a longitudinal field along $x$. {\bf For} the sake of clarity, we recall that in the case of the related current filamentation instability, where an effective anisotropy in velocity space arises from the presence of two counterstreaming electron beams \cite{CFI}, the growth of a longitudinal electric field  in a two dimensional  spatial configuration is due to the onset of a primary  two-stream instability with wave vector along the direction of the beams (i.e. along $y$ with the choice of axes considered here).
On the contrary,  in the present case,  the growth  of a longitudinal electric field  can only occur as a secondary process produced by the development of the Weibel instability itself.  We also note that the generation of a longitudinal field, orthogonal to the beam direction, during the nonlinear evolution of the current filamentation instability in a one dimensional configuration was discussed analytically\cite{Prandi, SPIKES} in terms of a two fluid, cold  plasma description and shown to occur near wave breaking conditions due to the nonlinear coupling between the current filamentation
instability and Langmuir waves.

In the present article we recover within a kinetic description the nonlinear generation  of a longitudinal field, orthogonal to
the direction  of the perturbed  current produced by  the  Weibel instability in its linear phase. Furthermore we show that the
nonlinear  time evolution of the electron distribution function in phase space, in the presence of the magnetic and inductive
electric fields produced by the Weibel instability, is characterized by a mixing process that occurs mostly at the position along $x$ where the  amplitude of the magnetic field is largest  and by particle acceleration or deceleration that occurs mostly  at the position along $x$ where the amplitude of the inductive electric  field is largest. The  combination of these processes  leads to the formation of a highly non monotonic distribution function in velocity space  that can further excite Langmuir waves resonantly.  Eventually,  the evolution  of the longitudinal  fields  is found to lead to the formation of multipolar electrostatic structures  characterized by a bi-polar or tri-polar electric field profile that are reminiscent of the "isolated electrostatic structures"  detected   by satellites {\em in-situ} in the solar wind plasma, in the auroral regions and in the   Earth magnetosphere and bow shock.  These structures have a typical width of the order of  a few tenth of a Debye length and are thought to  propagate along the magnetic field lines at speeds smaller than the solar wind speed. The tri-polar structures are also known as double-layers and are accompanied by a net potential drop. They have recently been proposed as a possible candidate in order to account, through  a sequence of such double layers, for the total potential difference from the solar corona to the Earth, as consistent with exospheric models (see the recent review \cite{sw} and references therein).

\section{Governing Equations}\label{GE}

We consider high frequency modes evolving on electron time scales {\bf defined by the inverse of plasma frequency} in  a collisionless  plasma and describe the dynamics of the
plasma  using the Vlasov-Maxwell system of equations for the electron  distribution function ${f}_{e}$ taking  the ions to
remain at rest.   Thus in dimensionless units  we write
\begin{equation} \frac{\partial{f}_{e}}{\partial{t}}+ \mathbf{v}.\frac{\partial{f}_{e}}{\partial{\mathbf{x}}}+
(\mathbf{E} + \mathbf{v} \times \mathbf{B}) \cdot\ \frac{\partial{f}_{e}}{\partial{\mathbf{v}}} = 0, \label{VLA}
\end{equation}
where  velocities are normalized to   the speed of light and times  to the inverse of the plasma frequency $\omega_{pe}$.
Lengths are thus normalized to the electron skin depth $d_e = c/\omega_{pe}$.

We restrict ourselves to  a 1D-2V configuration  where all quantities depend on $x$ and time only, the particle velocities and
the electric field have $x$-$y$ components  and the magnetic field is along the $z$ axis, and write Maxwell's equations in the
form
\begin{equation} \label{MAX}
\frac{\partial{B_{z}}}{\partial{t}} = -\frac{\partial{E_{y}}}{\partial{x}}, \qquad -\frac{\partial{B_{z}}}{\partial{x}} =
\frac{\partial{E_{y}}}{\partial{t}} + J_{y}, \qquad \frac{\partial^2\phi}{\partial^2x} = -\rho, \end{equation} where $\rho$ and
$\mathbf{j}$ are the dimensionless charge and current densities and $\phi$ the dimensionless electrostatic potential.

The Vlasov equation (\ref{VLA}) is integrated in the phase space $(x,v_{x},v_{y})$ with periodic boundary conditions along the
$x$ direction.  For a description of the adopted code see Ref. \cite{Mang}. We take  the following initial condition
\begin{equation} f_{e}(x,v_{x},v_{y},t=0) = f_{M}(v_{x},v_{y})[1 + \epsilon \sum^{N}_{n=1}\ \cos(k_nx+\phi_n))]; \hspace{5mm}
k_{n}=2\pi n/L_{x}.
\label{IC1}
\end{equation}
Here $L_{x}$ is the length of the simulation domain in the $x$ direction, $\epsilon$ is the
perturbation amplitude varied in the range $10^{-6} \leq \epsilon \leq 10^{-4}$ and $f_{M}(v)$ is a two-temperature Maxwellian distribution given by
\begin{equation} f_{M}(v_{x},v_{y}) = \frac{n_e}{\pi \sqrt{T_{x}T_{y}}}\\
exp\Big(-\frac{v_{x}^2}{T_{x}}-\frac{v_{y}^2}{T_{y}}\Big),
\end{equation}
with $T_{x}= 4\cdot 10^{-4}$ corresponding to  a
plasma with a non-relativistic temperature,  $T_{y}/T_{x} = 12$ and $\phi_n$ are random phases.  At $t=0$ we  also introduce a perturbation on the magnetic field:
\begin{equation}
B_{z} = a_{i} \sum^{N}_{n=1}\ [\cos(k_{n}x+\psi_n)],
\label{IC2}
\end{equation}
{\bf where $a_{i}$} is the initial field amplitude for each wave and $\psi_n$ are random phases. We take $a_{i} = 10^{-4}$ and $N=200$ in both Eqs. (\ref{IC1}) and (\ref{IC2}). We note the initial conditions with random initial amplitudes of the density and magnetic and electric field have been also used without any significative qualitative and quantitative change in the results.

We consider a field free plasma, hence $<B_{z}> = 0$ at $t= 0$. {\bf The computational velocity phase space is given by $[-v_x^{max}, v_x^{max}] ~ \times ~ [-v_y^{max}, v_y^{max}]$. We take $v_{x}^{max} = 0.1$ and  $v_{y}^{max} = 0.27$ since the system is characterized by an initial  temperature anisotropy $T_y > T_x$. However, the distribution function  will spontaneously evolve towards a more isotropic configuration  by reducing its width in the $v_y$ direction and  by generating  small scale filamentary structures with the same  characteristic size in both directions. Therefore, we take a larger number of points in the $v_y$ direction in order to reduce the difference between the resolution in the two directions. For computational reasons (i.e.  not  to increase  the total number of grid points too much), a good compromise is  $N_{v_{x}} = 281$ and  $N_{v_{y}} = 121$. 
The time step is $\Delta t = 0.005  \omega_{pe}^{-1}$. The evolution of the system is investigated up to $t = 500\, \omega_{pe}^{-1}$. 
 The length of the spatial simulation  domain is $L_x= 6\pi d_{e}$. The space discretization along  $x$ is $dx = 0.03 \, d_{e}$,  where $d_{e}$ is the electron skin depth, corresponding to $N_{x} = 600$ grid points in the   $x$ direction. This allows us to account
for phenomena that occur on small spatial scales that are  larger than, but of the order of, the electron Debye scale $\lambda_D = 0.02 \,d_e$. }

\section{Onset of the Weibel instability and early nonlinear phase}\label{onset}

\begin{figure}[!t]
\centerline{\psfig{figure=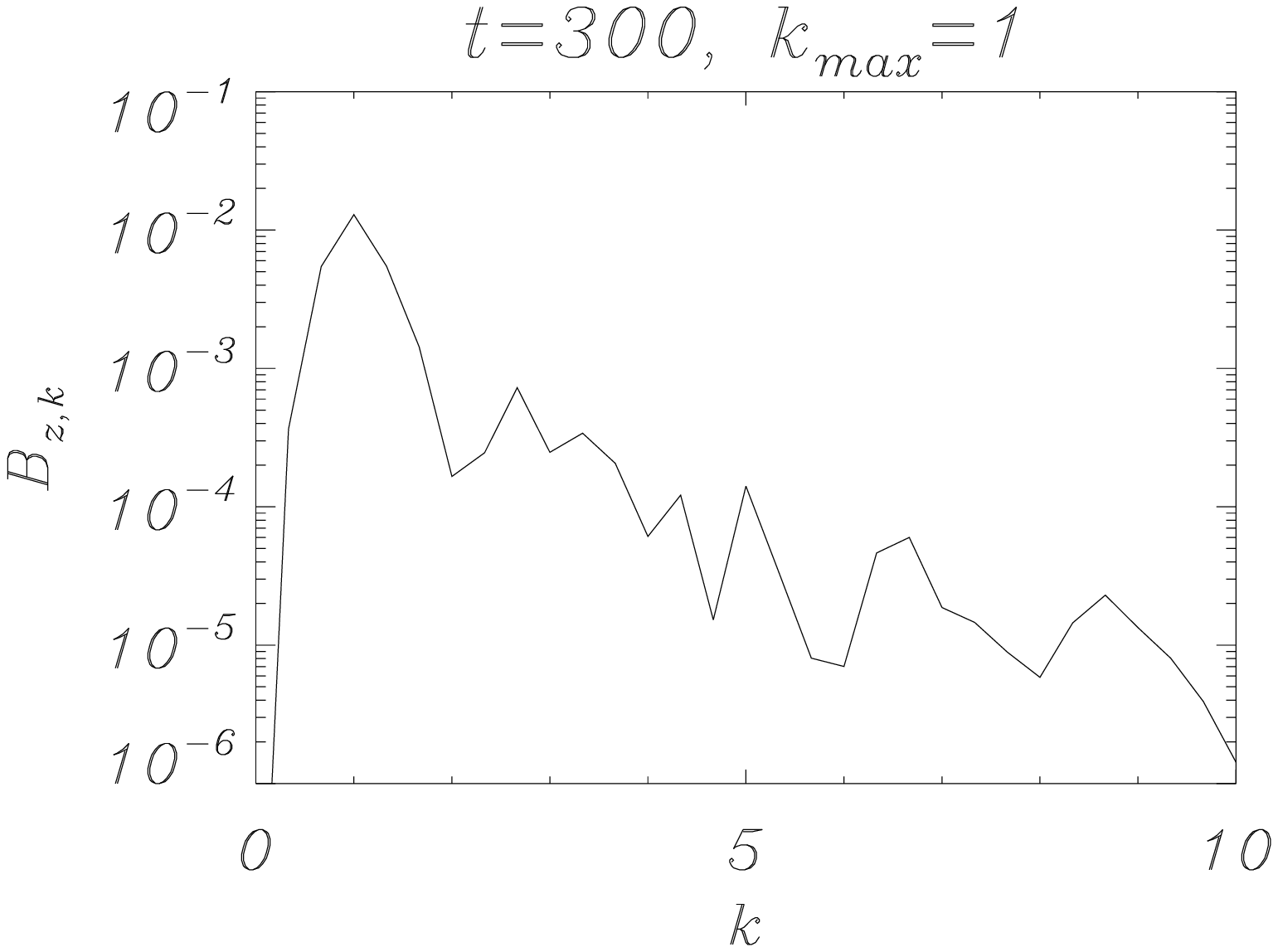,height=7cm,width=8cm}\psfig{figure=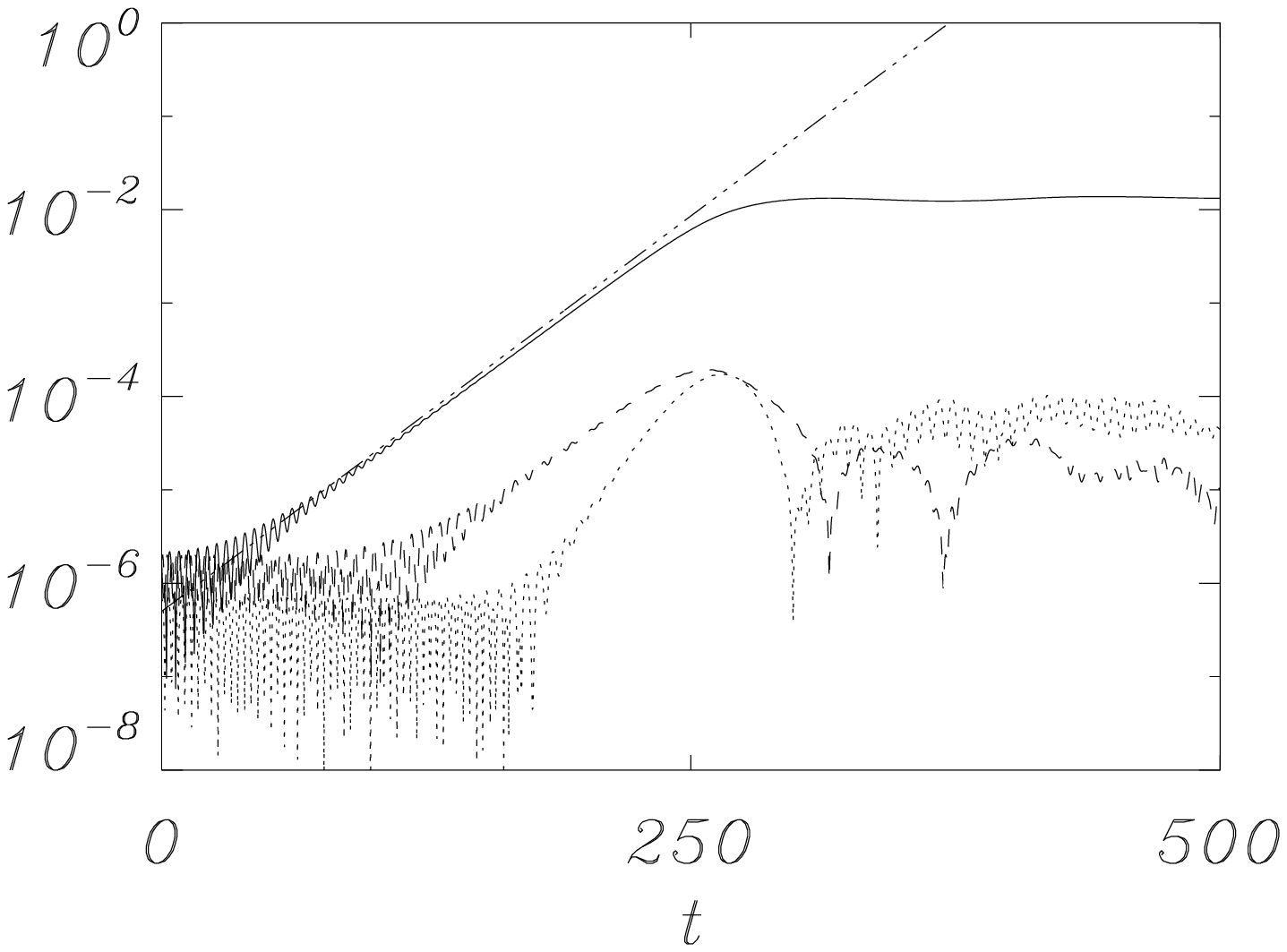,height=7cm,width=8.cm}}
\caption[ ]{\small Left frame: amplitude of the Fourier components of the magnetic field  $B_{z,k}$ at $t = 300$  vs. $k$. Right frame: time evolution of the Fourier components of the most unstable $k=k_{max}=1$ magnetic and inductive electric field mode, $B_{z,k=1}$ and $E_{y,k=1}$, solid and dashed line respectively, and of the most unstable $k=2.3$ longitudinal electric field $E_{x,k=2.3}$ (dotted line). The dash-three dotted line represents the growth rate of the magnetic field.}
\label{Fig1}
\end{figure}

\begin{figure}[!t]
\centerline{\psfig{figure=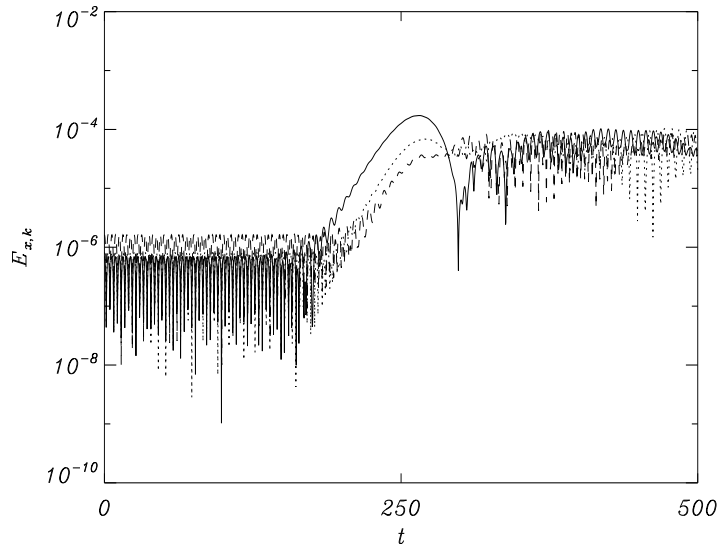,height=8cm,width=10cm}}
 \caption[ ]{\small Growth of $k=1, 2, 2.3$ Fourier components of the longitudinal field $E_{x,k}$, dashed, dotted {\bf and } solid line, respectively.}
\label{Fig2}
\end{figure}

The onset and the linear growth of the Weibel instability  up to the beginning of its saturation phase is shown in Fig. \ref{Fig1},  as obtained from the numerical integration of the Vlasov equation described above. In the left frame the Fourier amplitude of the magnetic field $B_{z,k}$  at $t = 300$  is shown  vs. $k$ (note that  only a part of the interval in $k$ is shown). We see that the modes in the range $0.3 \le k \le 1.5$ have grown significantly with their maximum growth rate corresponding to $k = 1$, {\bf  i.e. in dimensional units to  $k = d_e^{-1}$}. In the right frame,  the amplitude of the magnetic field $B_{z,k}$ vs. time  is plotted for the most unstable mode, $k = 1$ (solid line).  After the initial transient phase the amplitude of the magnetic field grows exponentially, for $t \le 270$ (linear phase),  with growth rate $\gamma_{max} \sim 0.04$ (see dash-three dotted line),  consistent with the analytical results (see  Ref. \cite{WEIBEL}).

The growth of the  inductive electric field component  $E_{y,k}$ associated to $B_{z,k}$  is shown in the same frame (dashed line) still  for $k = 1$ , together  with  the longitudinal field $E_{x,k}$ (dotted line) taken at  $k =2.3$. We note the $E_{y,k}$ reaches its  maximum value at the saturation time of  $B_{z,k}$ and then starts  to decrease. This is consistent  with  the fact that  the ratio between the inductive electric field and the magnetic field generated by the Weibel instability scales  with the instability growth rate  so that,  as soon as the Weibel instability saturates, the  mode  structure is dominated by the magnetic field. In this phase the longitudinal  electric  field $E_{x,k}$ arises  because of the coupling between the Weibel instability and the Langmuir waves, as  discussed within the fluid approximation in Ref. \cite{SPIKES},  due to the electron density modulation induced by the spatial modulation of $B^2_{z,k}$, and thus grows in time  at twice the  growth rate of the magnetic field.

While the  Fourier component  of the longitudinal  field shown in Fig. \ref{Fig1}, right frame,  corresponds  to $k=2.3$,   the spectrum  in $k$ of the longitudinal field and of the density modulation  is wider, as consistent  with the fact that the Weibel instability grows with very close growth rates in a relatively large  range of values of $k$ around $k\sim 1$.   As shown in Fig. \ref{Fig2}, different Fourier  components of the longitudinal field  grow with nearly equal values of the growth rate. As a consequence, while the longitudinal field grows two times as fast as the magnetic field, its spatial structure contains additional wavelengths besides $\lambda/2= \pi$.

\begin{figure}[!t]
\centerline{\psfig{figure=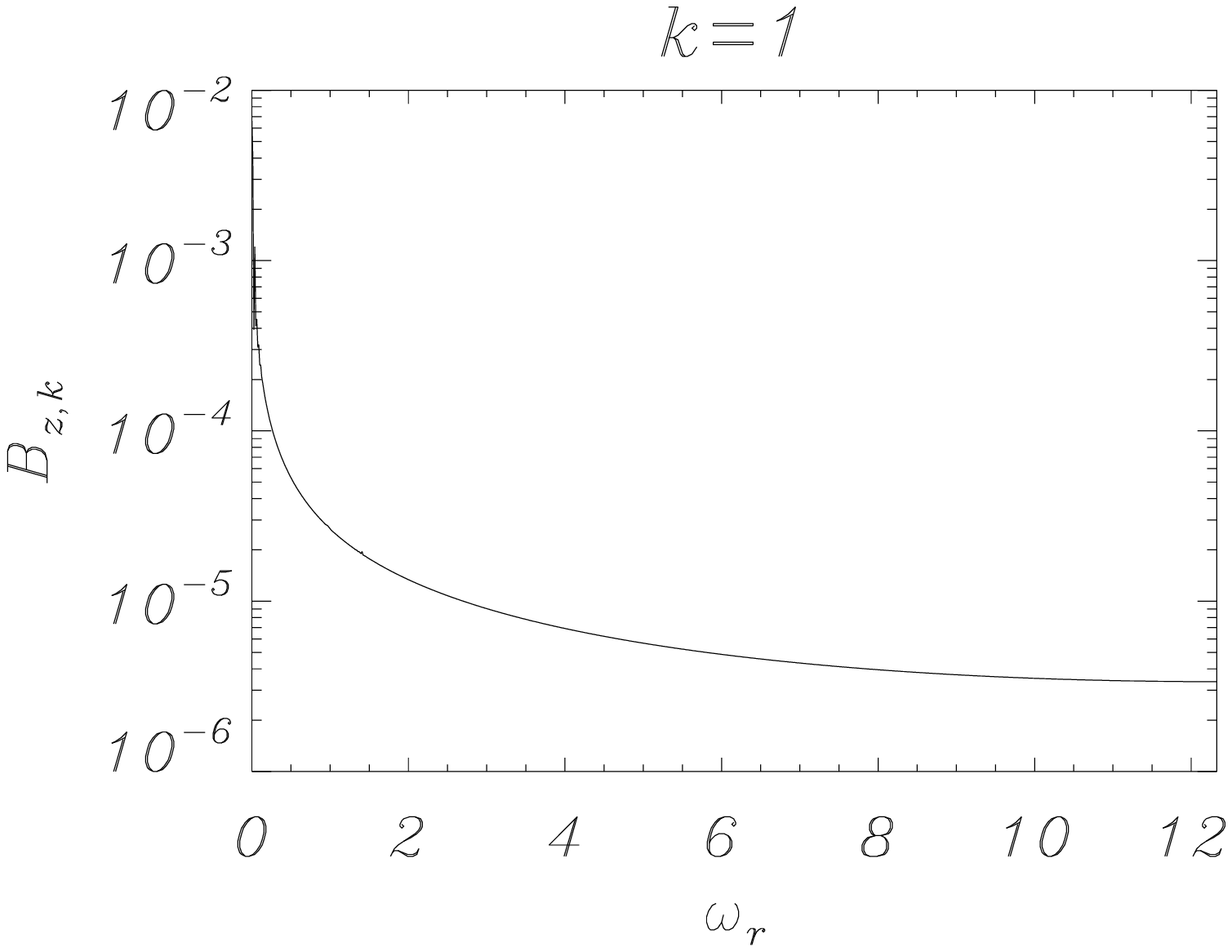,height=5cm,width=6cm}\psfig{figure=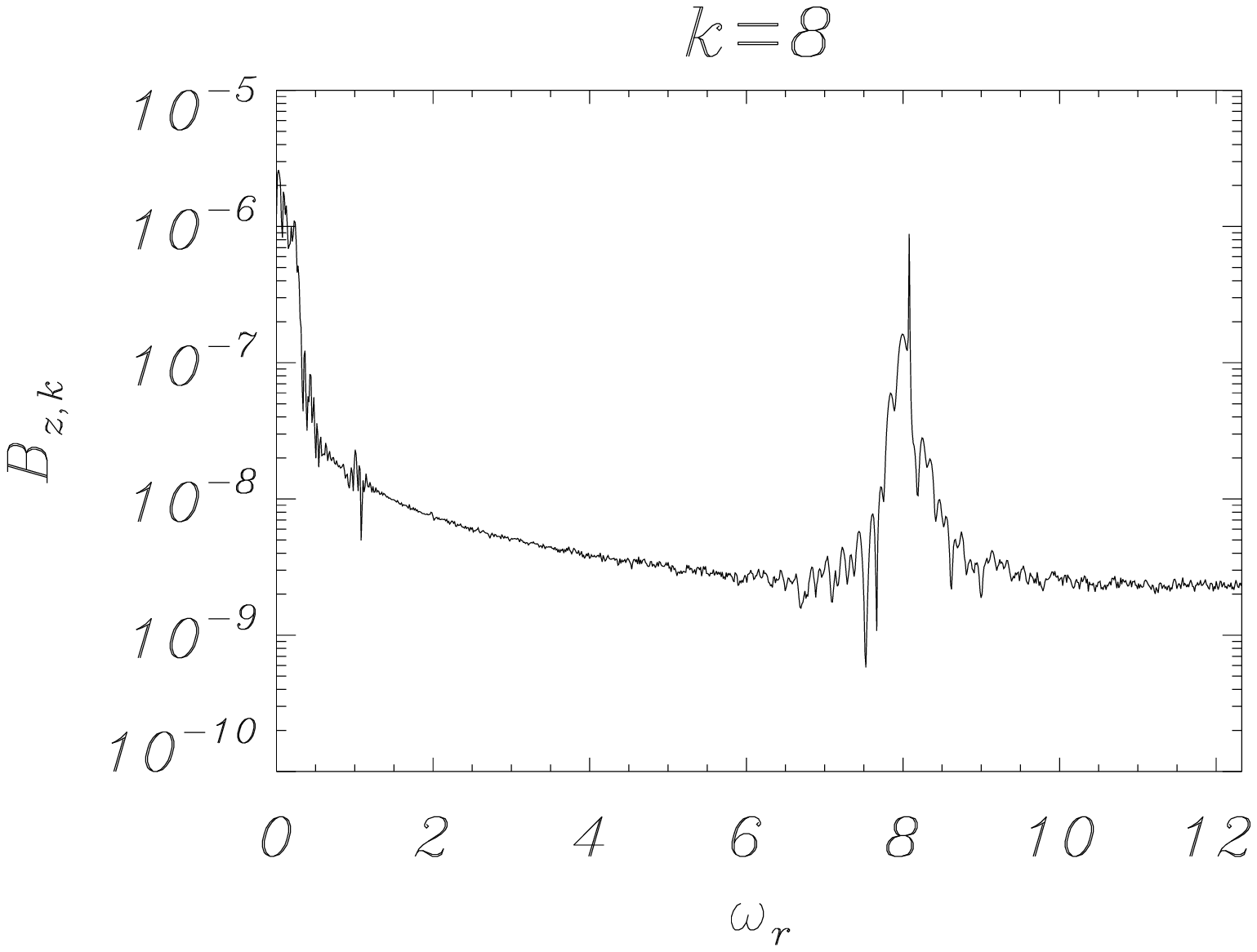,height=5cm,width=6cm}\psfig{figure=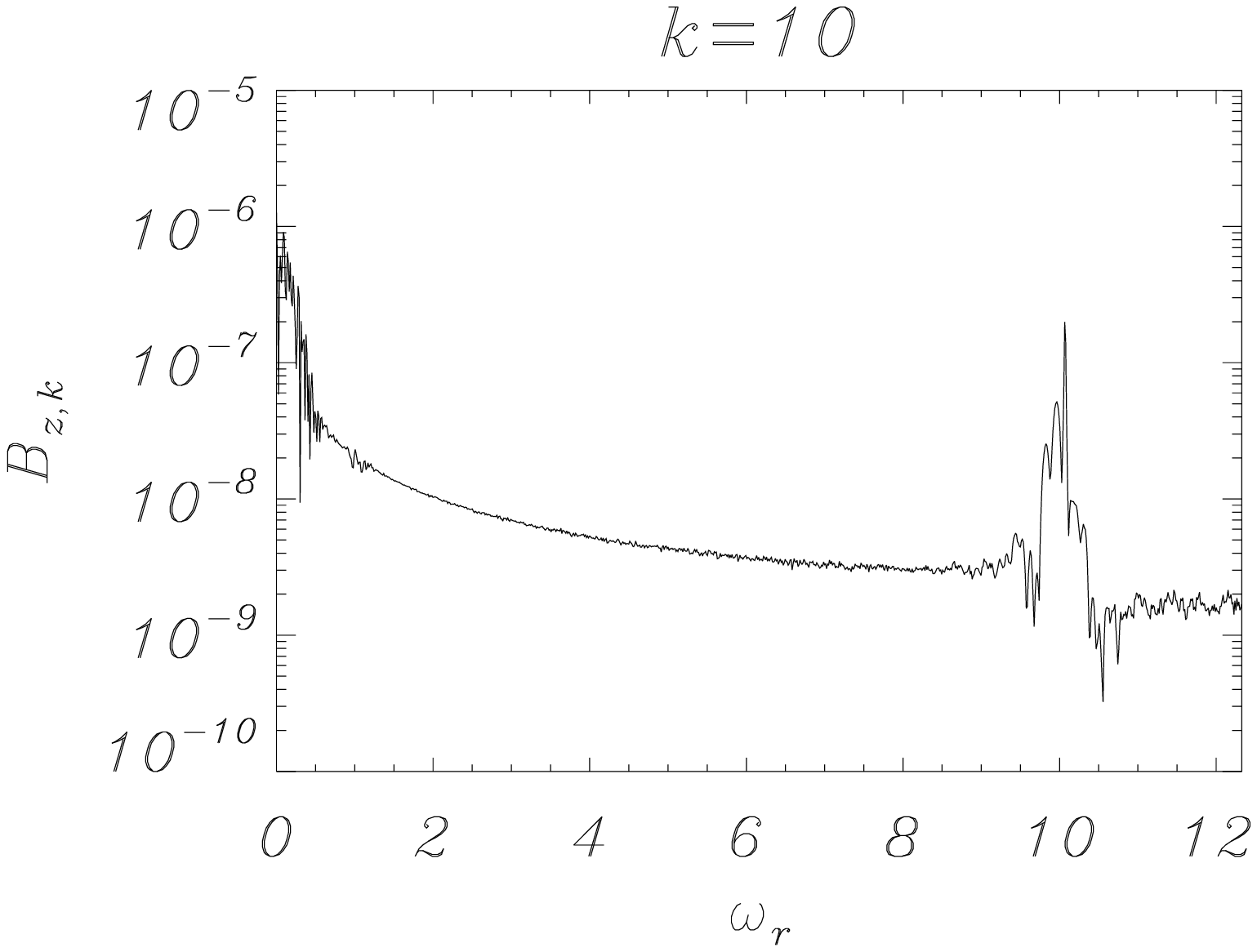,height=5cm,width=6cm}}
\caption[ ]{\small  Frequency spectra of  the magnetic field component $B_{z,k}$  for  $k = 1, 8 , 10$ (from left to right).}
\label{Fig3}
\end{figure}

The  frequency spectra of $B_{z,k}$ for fixed $k$,  are shown in Fig. \ref{Fig3} for $k=1, 8, 10$.  The spectral amplitude  is  largest for $k =1$  and in the low frequency part of the frequency spectrum corresponding to the Weibel instability. For larger values of $k$  the frequency spectra   exhibit  a weaker feature  for large frequencies  corresponding to low amplitude transverse electromagnetic waves  propagating essentially at the speed of light ($\omega \sim k$ in dimensionless units).

\begin{figure}[!t]
\centerline{\psfig{figure=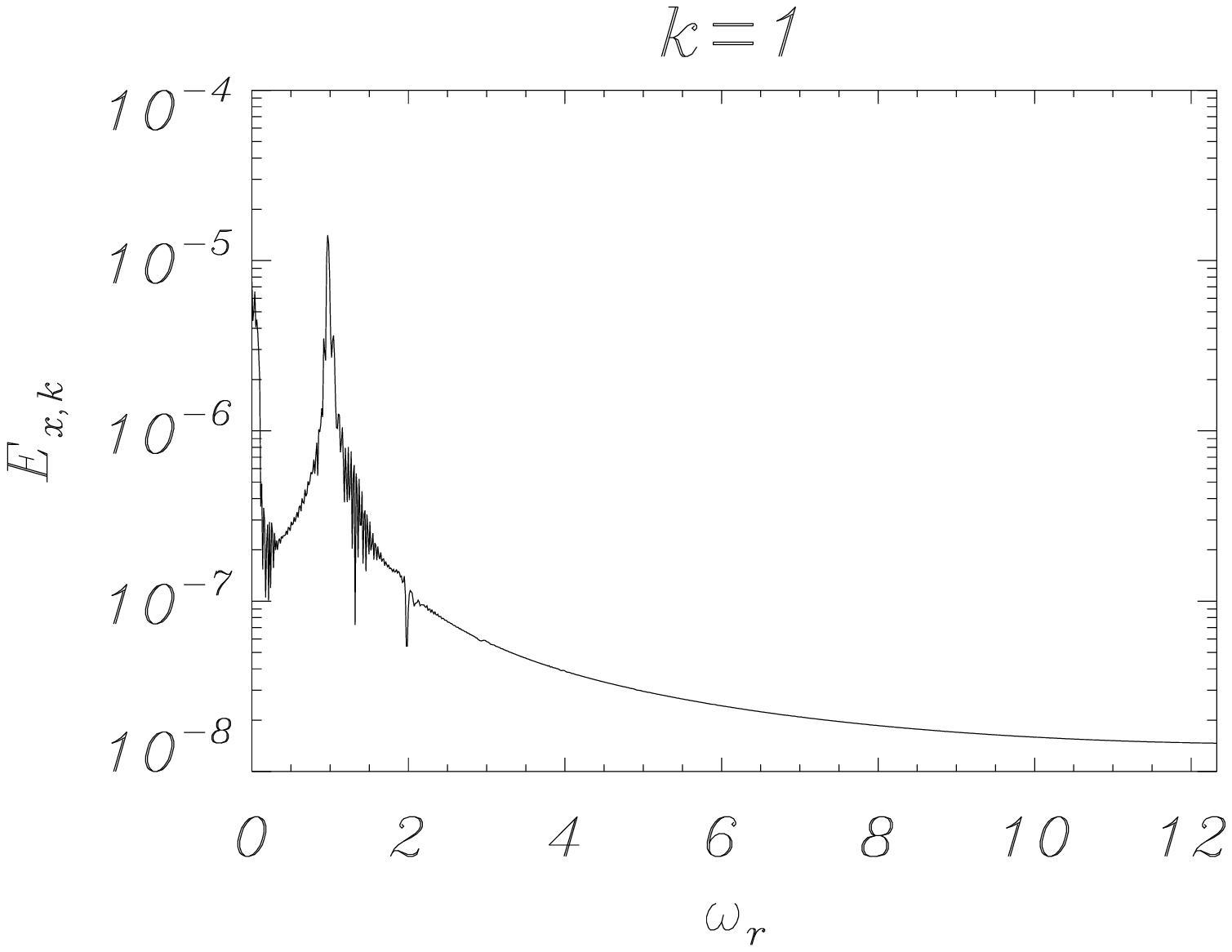,height=5cm,width=6cm}\psfig{figure=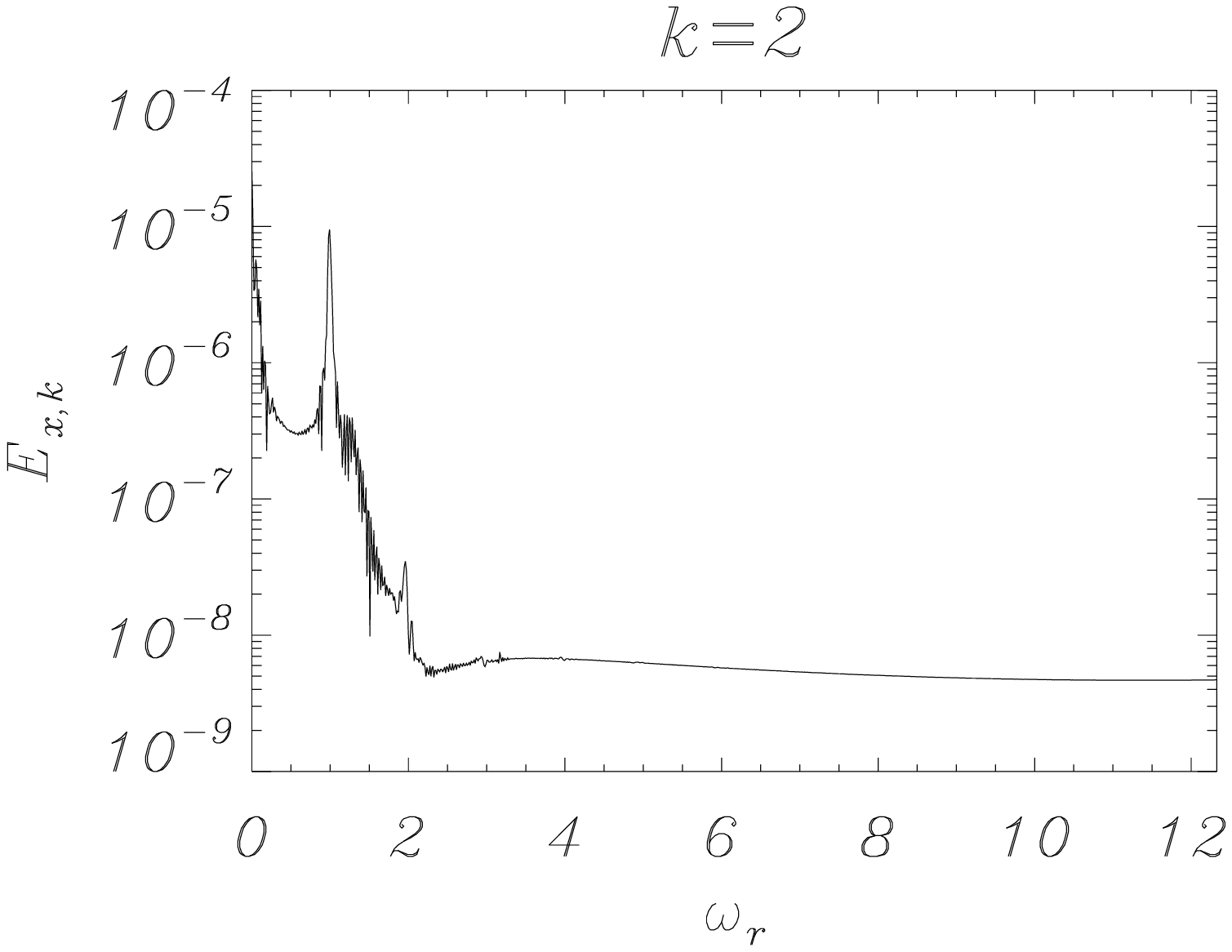,height=5cm,width=6cm}\psfig{figure=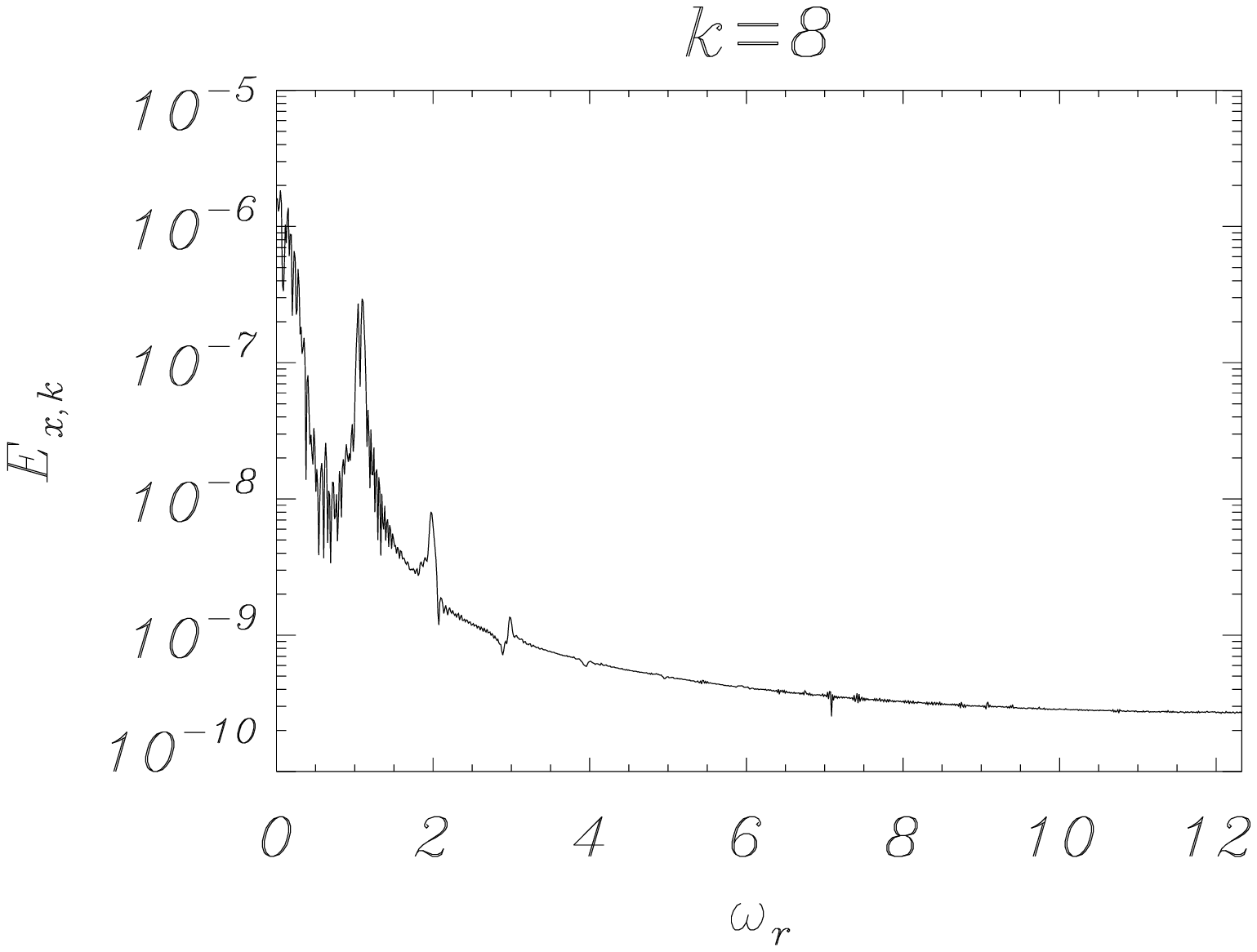,height=5cm,width=6cm}}
\caption[ ]{\small Frequency spectra of the longitudinal electric field component $E_{x,k}$ for $k = 1, 2, 8$ (from left to right).}
\label{Fig4}
\end{figure}

The  frequency spectra of the longitudinal  field $E_{x,k}$ at fixed $k$,  are shown in Fig. \ref{Fig4} for $k=1,  2,  8$ and  exhibit  a  clear peak at the electron  plasma frequency. This peak appears to be  less pronounced  for larger values of $k$.

{\bf Additional numerical integrations of the Vlasov equation, not  shown here,   for different initial anisotropy ratios ($T_y/T_x = 9$  and $T_y/T_x = 4$)  at constant $T_x$ indicate  that, while the $k=1$ mode remains the most unstable electromagnetic mode, the spectrum of the excited modes   becomes narrower with decreasing anisotropy,  since  higher $k$  modes grow less fast, as consistent   with linear theory.  This effect is also present in the spectrum of the longitudinal electric field $E_{x,k}$ in which case the value of $k$ corresponding to the maximum growth rate decreases as the anisotropy ratio is decreased.   The maximum value of $B_z$, taken  at the time when its growth is saturated,  decreases with decreasing anisotropy ratio.  Conversely, if $T_x$ is decreased at constant anisotropy ratio ($T_y/T_x =12$),  the maximum value of $B_z$ is almost unchanged,  while the value of $k$ corresponding to the maximum growth rate of the longitudinal electric field decreases from  $k=2.3$ at $T_x = 1$ to  $k=2$ at $T_x = 0.5$. }

\section{Phase space evolution in the advanced nonlinear phase}\label{NL}

\begin{figure}[!t]
\centerline{\hspace{1cm} \psfig{figure=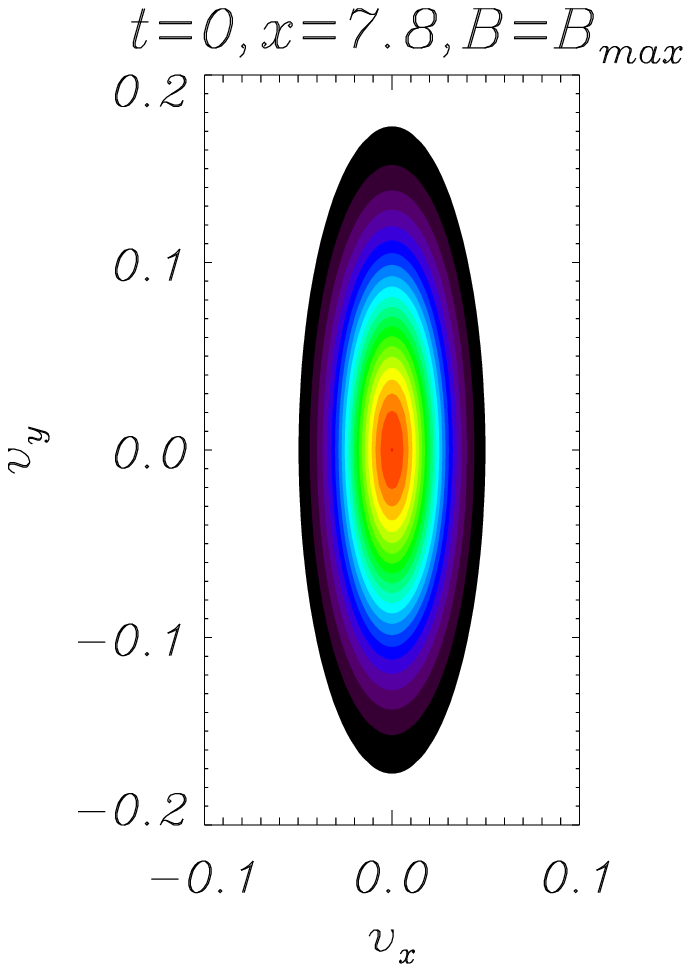,height=5cm,width=5cm}
\hspace{-3cm}\psfig{figure=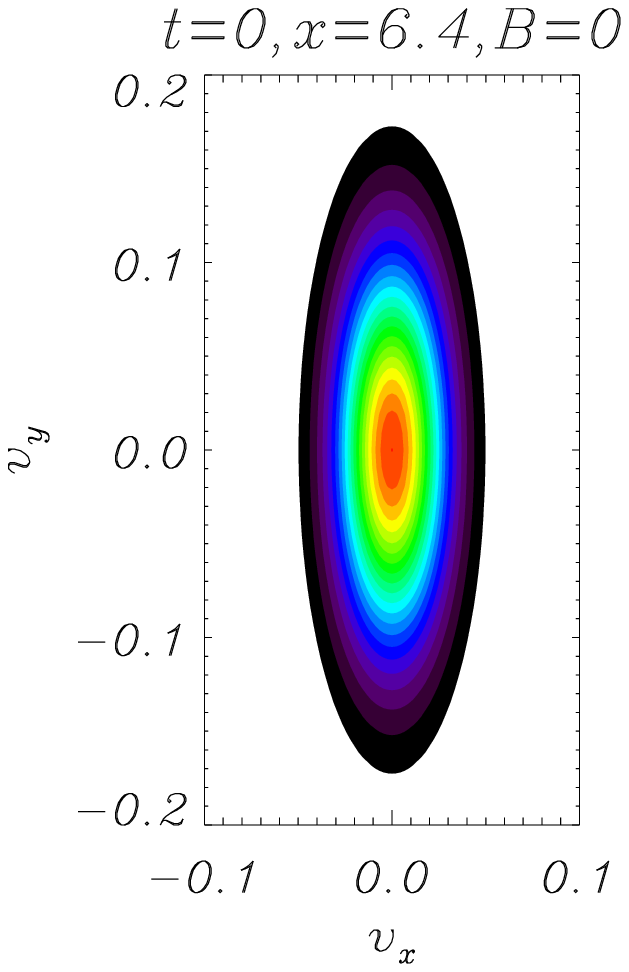,height=5cm,width=5cm}\hspace{-3cm}
\psfig{figure=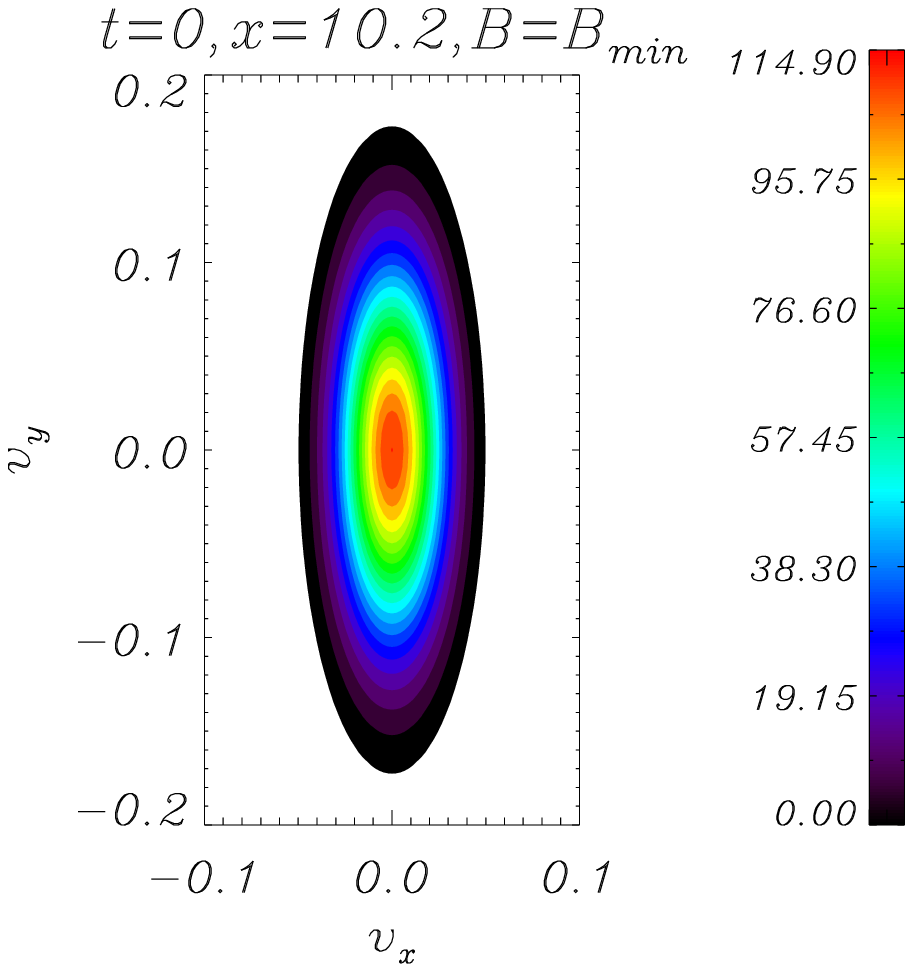,height=5cm,width=5cm}}
\centerline{\hspace{1cm} \psfig{figure=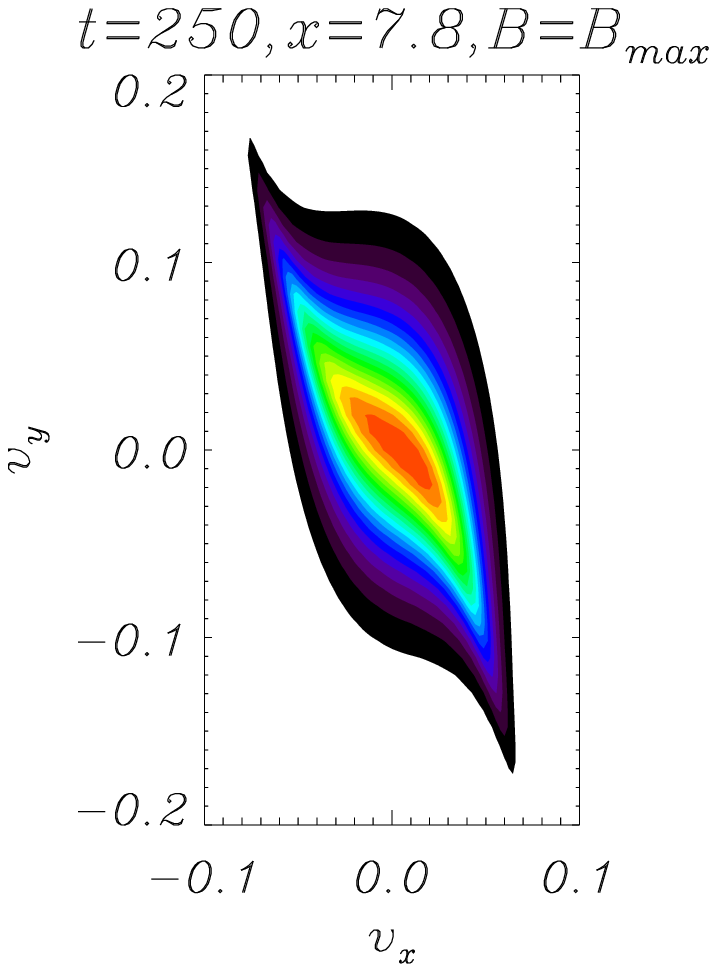,height=5cm,width=5cm}
\hspace{-3cm}\psfig{figure=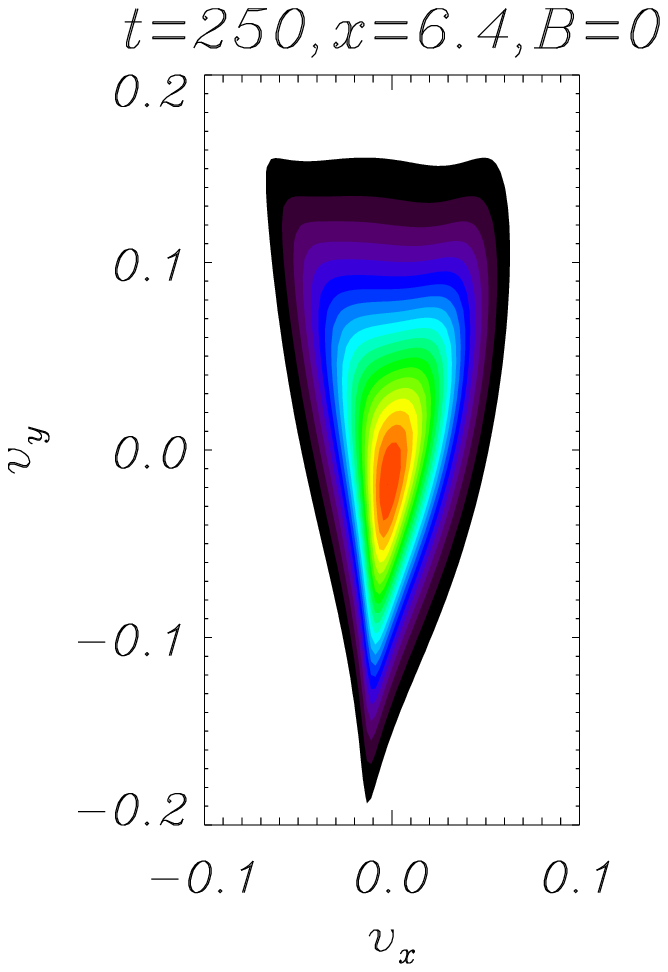,height=5cm,width=5cm}\hspace{-3cm}
\psfig{figure=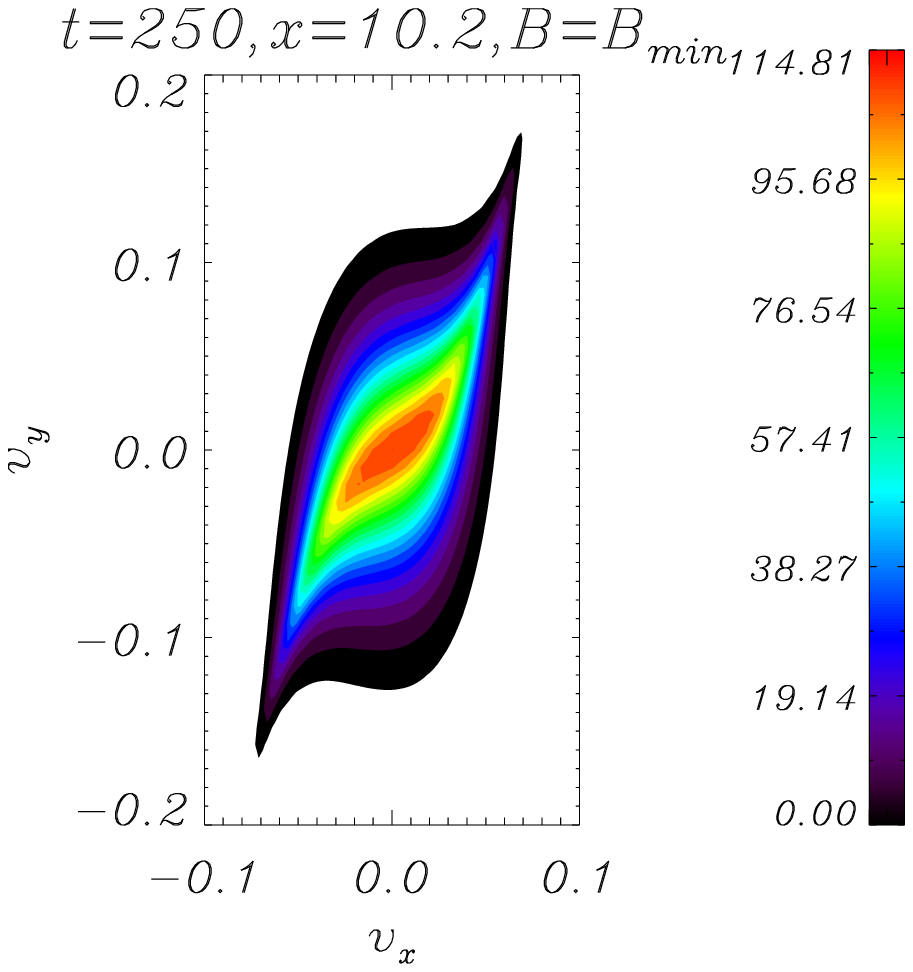,height=5cm,width=5cm}}
\centerline{\hspace{1cm} \psfig{figure=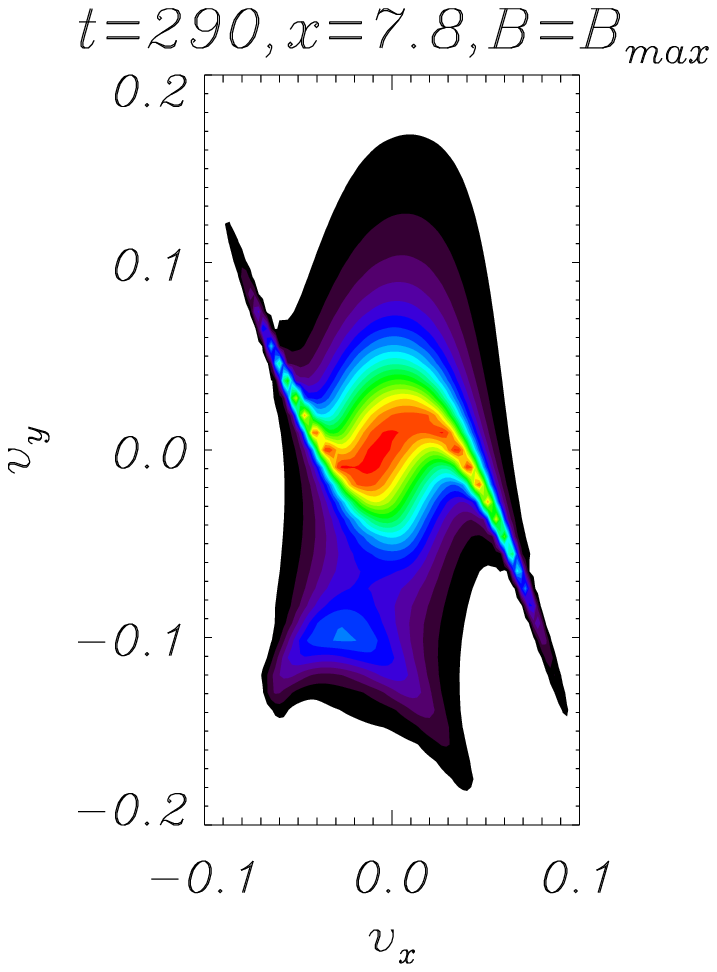,height=5cm,width=5cm}
\hspace{-3cm}\psfig{figure=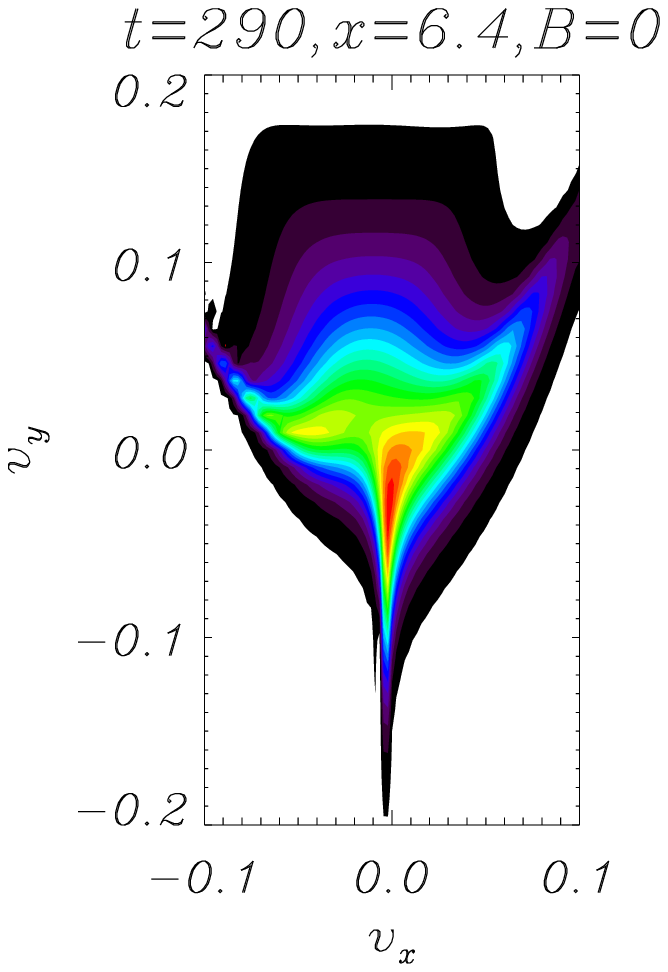,height=5cm,width=5cm}\hspace{-3cm}
\psfig{figure=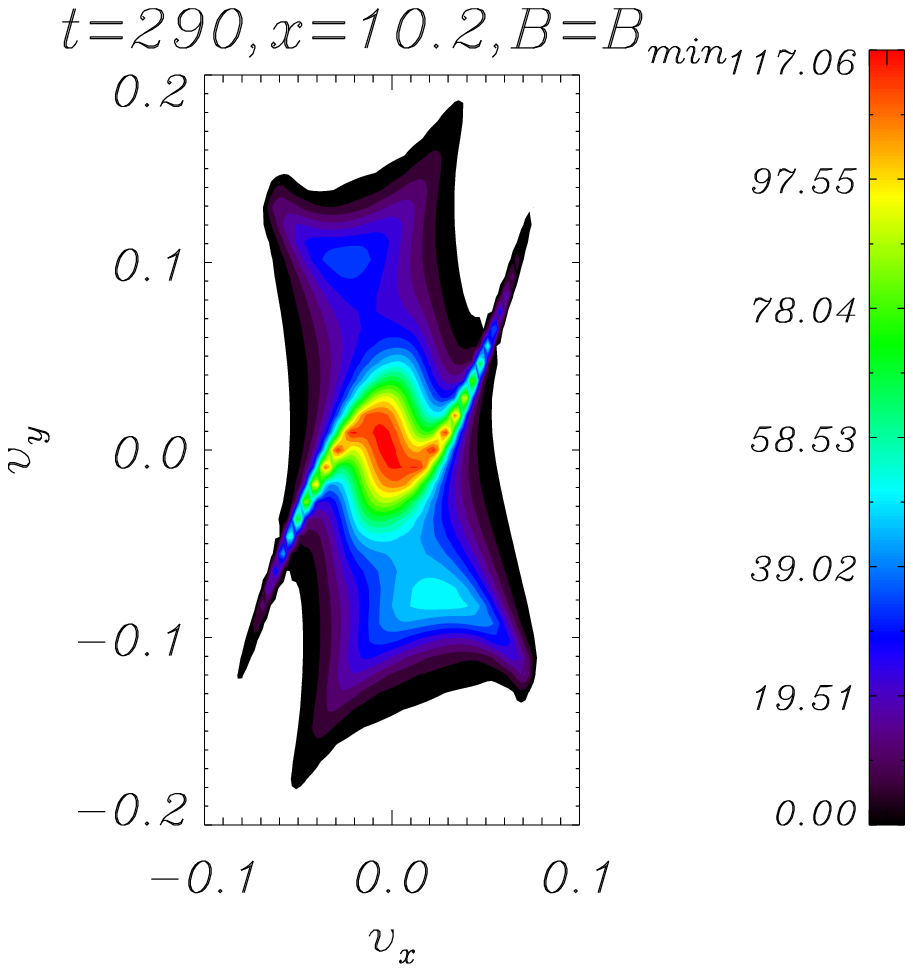,height=5cm,width=5cm}}
\centerline{\hspace{1cm} \psfig{figure=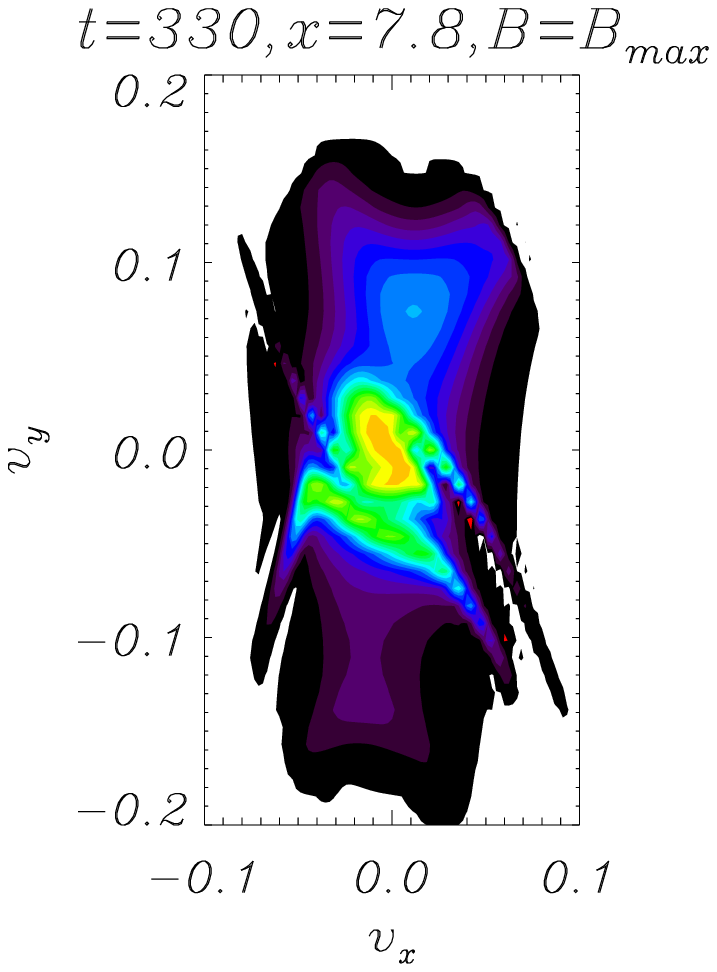,height=5cm,width=5cm}
\hspace{-3cm}\psfig{figure=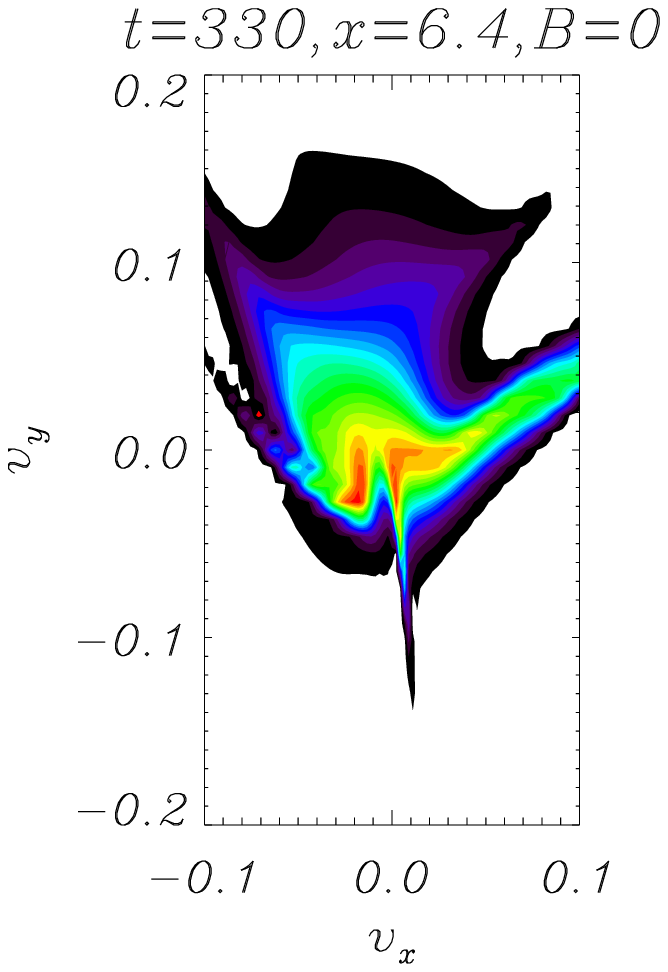,height=5cm,width=5cm}\hspace{-3cm}
\psfig{figure=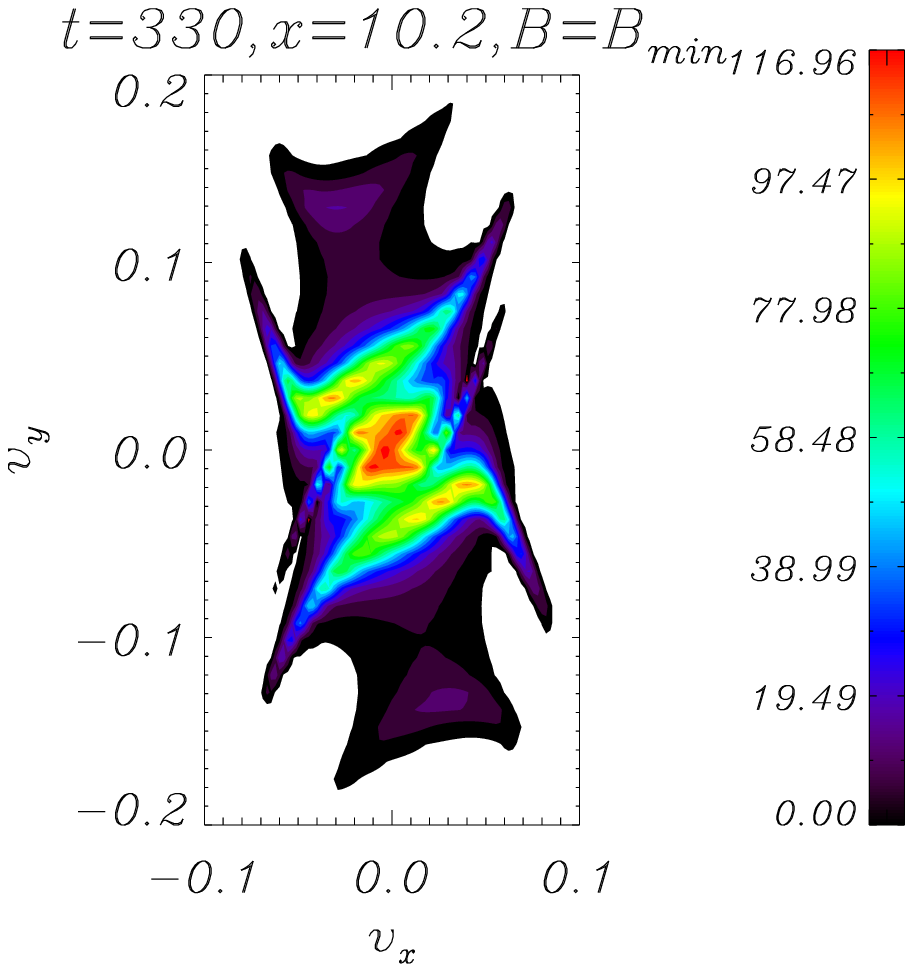,height=5cm,width=5cm}}
 \caption[ ]{\small  {\bf Time evolution of the electron distribution function with $t=0$(first row), $t=250$(second row), $t=290$(third row) and $t=330$(last row) at different positions along $x=7.8, 6.4, 10.2$ (from left to right)}. {\bf  Red (blue)  part is  most (least) densely populated.} }
\label{Fig5}
\end{figure}

\begin{figure}[!t]
\centerline{\psfig{figure=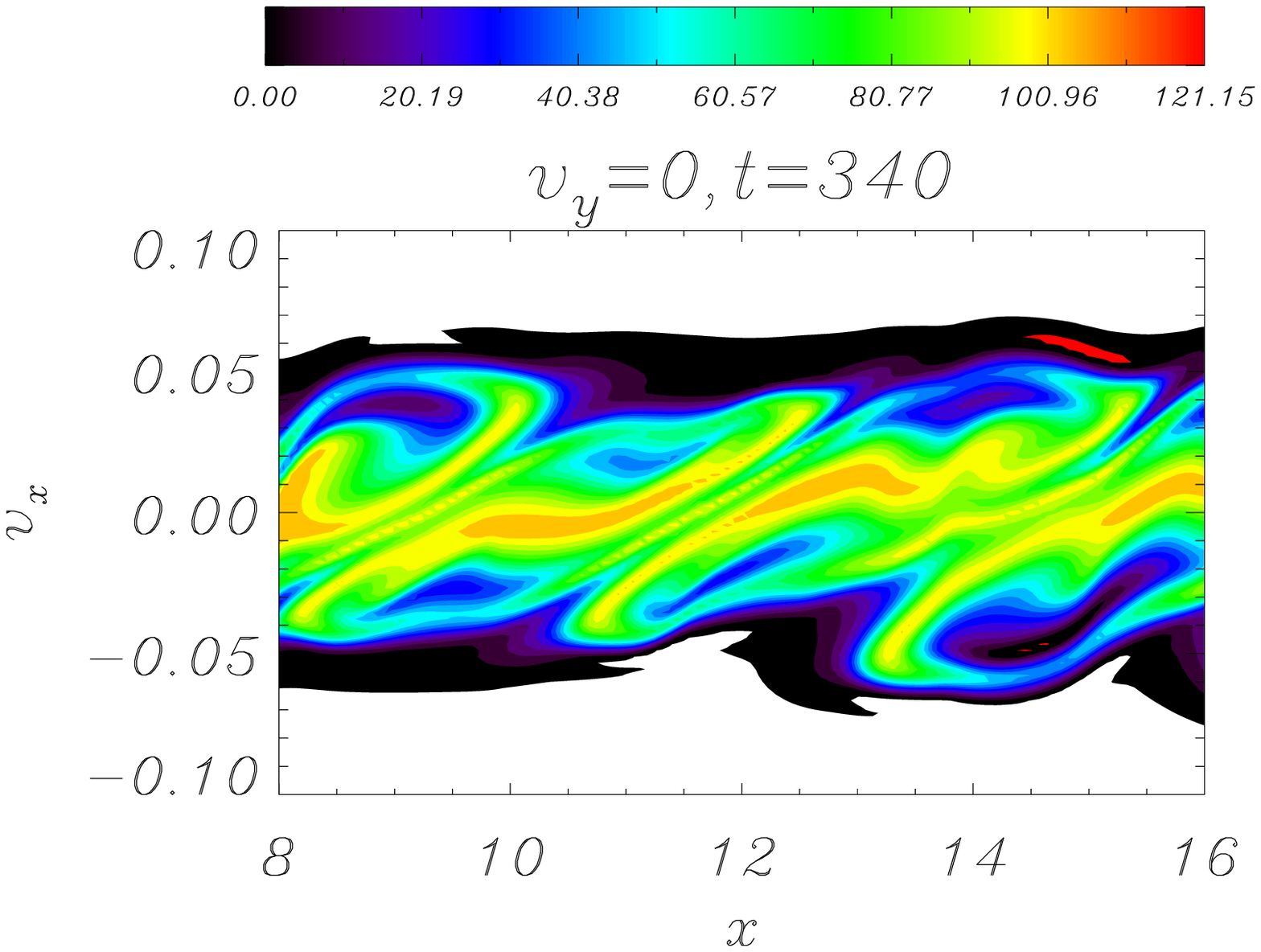,height=5cm,width=5cm}
\psfig{figure=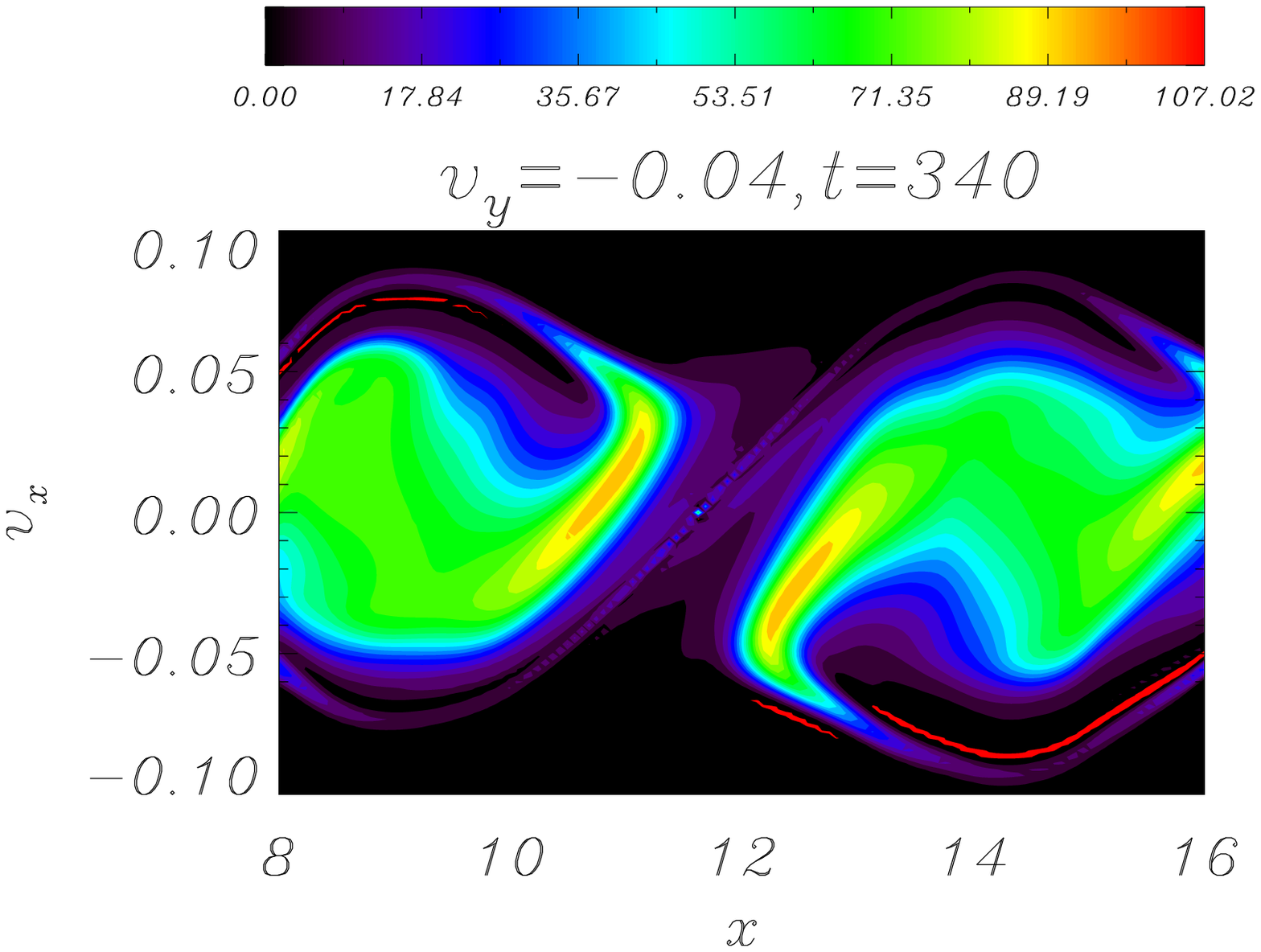,height=5cm,width=5cm}}
\centerline{\psfig{figure=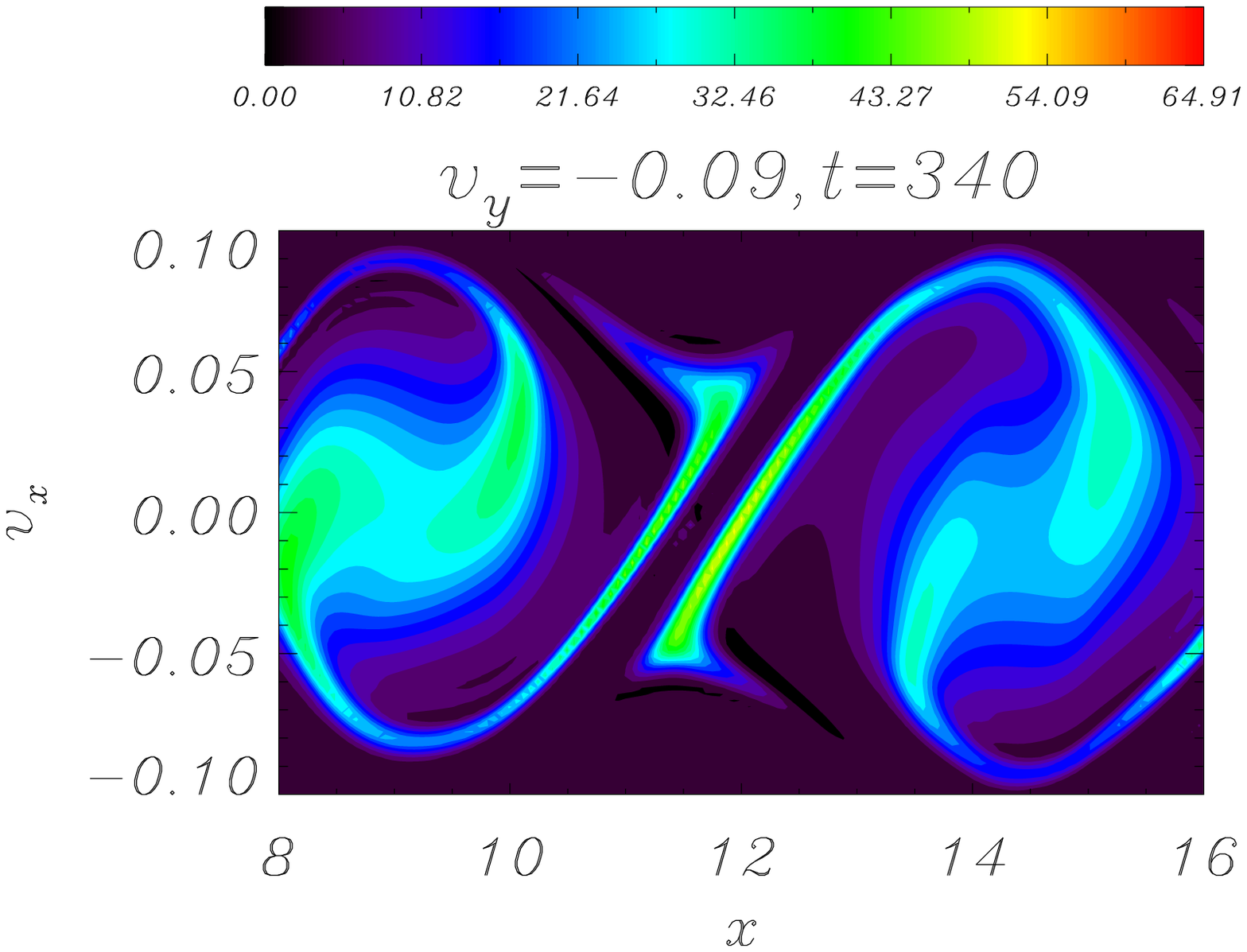,height=5cm,width=5cm}
\psfig{figure=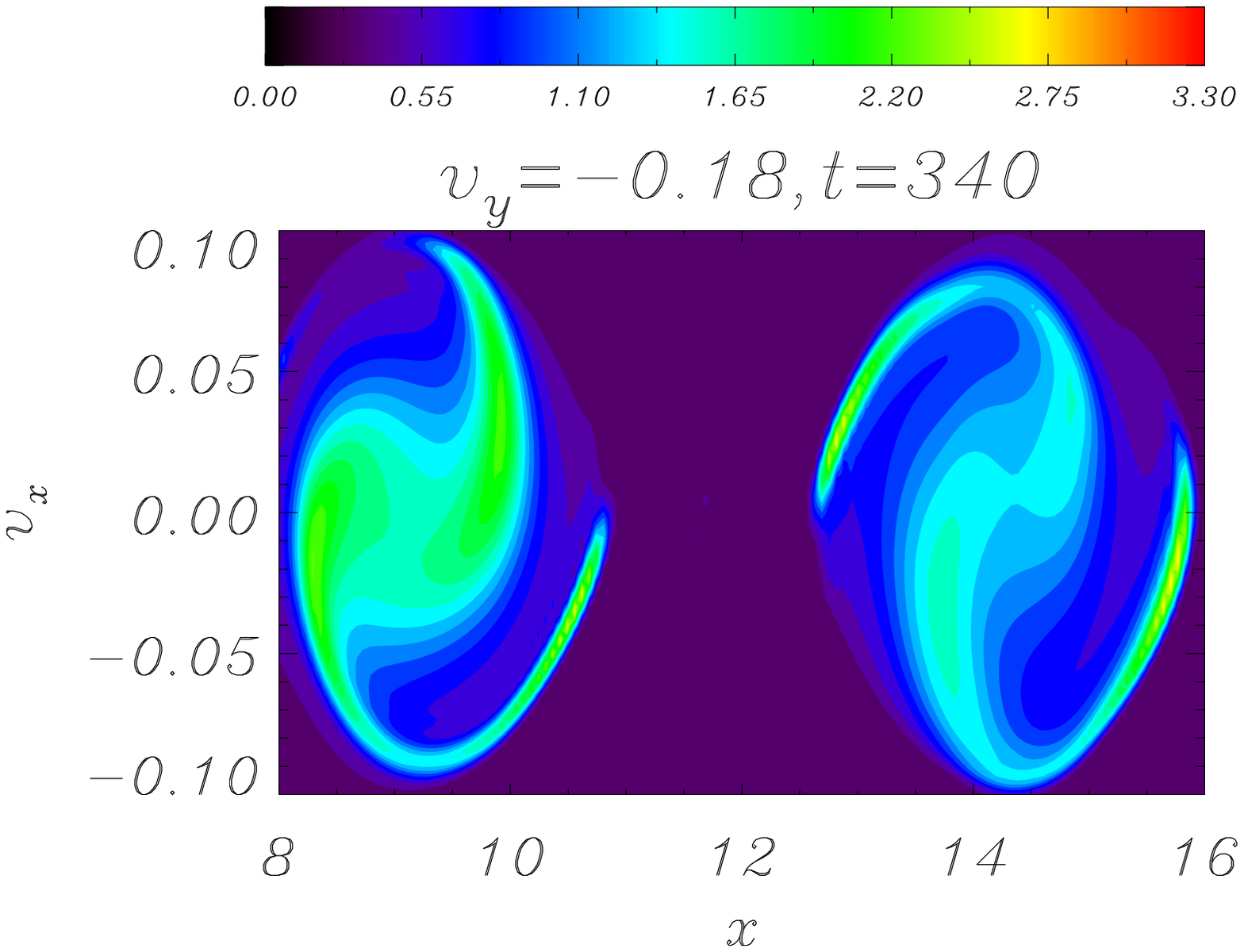,height=5cm,width=5cm}}
\caption[ ]{\small  Contour plots of the distribution function in the $x$-$v_x$ space  at $t = 340$  for different values of $v_y$.}
\label{Fig6}
\end{figure}

During the linear growth  of the instability,  the evolution of the electron distribution function in velocity space is characterized  by a differential rotation in velocity space at the points where the absolute value $|B_z|$ of the magnetic field   generated by the Weibel instability  has a maximum and  by a Y- shaped deformation  with axis along $v_y$  at the points where the perturbed magnetic field vanishes  which correspond to the points where the absolute value $|E_y|$  of the inductive electric field  is largest. This behaviour is shown by the frames in the second row  of Fig. \ref{Fig5}.  The rotation of the distribution function is clockwise or anti-clockwise depending on the sign of $B_z$ (labelled as $B_{max}$ and $B_{min}$ respectively in Fig. \ref{Fig5},  while the Y-shaped deformation is turned upside down depending on the sign of  $E_y$ (only one sign is shown in Fig. \ref{Fig5}).  As the Weibel instability enters its fully nonlinear phase the winding of the distribution function becomes tighter and the Y- deformation more marked until both features become ''multi-armed'', as shown by the last row in Fig. \ref{Fig5}.
The winding of the distribution function  and the formation of phase space vortices  associated to the magnetic field $B_z(x)$ is also apparent  from the contour plots of the distribution function   at $t = 340$ in the $x$-$v_x$ space shown in Fig. \ref{Fig6} for different values of $v_y$.

\begin{figure}[!t]
\centerline{\hspace{2cm} \psfig{figure= 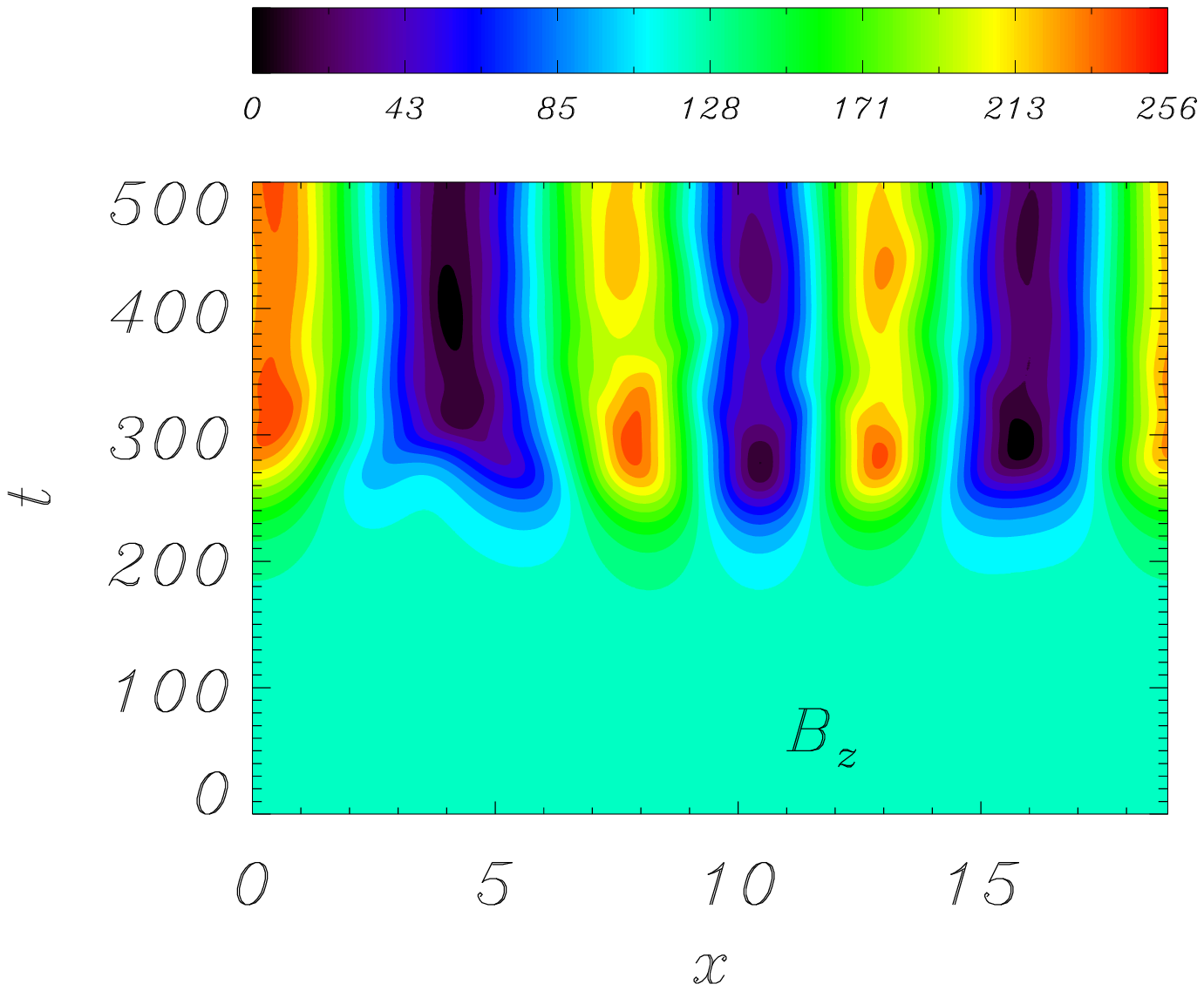,height=7cm,width=9cm}\hspace{-2cm}
\psfig{figure= 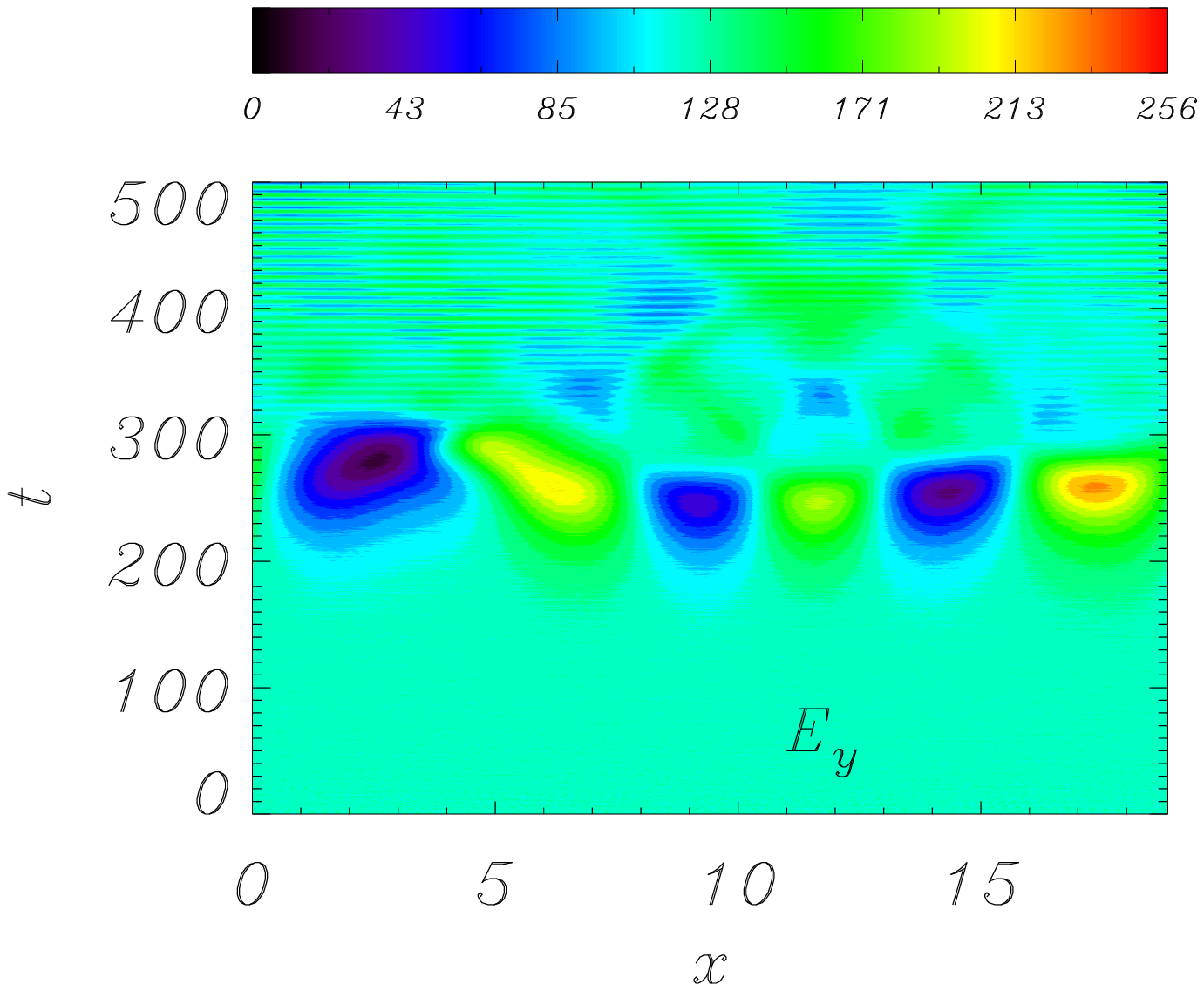,height=7cm,width=9cm}}
\caption[ ]{\small Contour plots showing the combined $x$-$t$ dependence of $B_z$  (left frame)  and of  $E_y$ (right frame).}
\label{Fig7}
\end{figure}

Disregarding for the moment the effect of the longitudinal electric  field, this complex behaviour can be attributed to the combined effects of the magnetic field  $B_z$,  that deflects the electron orbits (where $|B_z|$ is largest and $E_y$ vanishes) and of the  inductive electric field $E_y$ that accelerates or decelerates the electrons along $y$ (where $E_y$  is largest and $|B_z|$  vanishes). In order to visualize  this combined effect of the instability fields  on the plasma electrons a set of particle orbits have been computed by  direct integration of the particle equation of motion in the electromagnetic fields
obtained from the Vlasov integration (shown in  Fig. \ref{Fig7} in ($x, t$) space).

\begin{figure}[!t]
\centerline{\psfig{figure=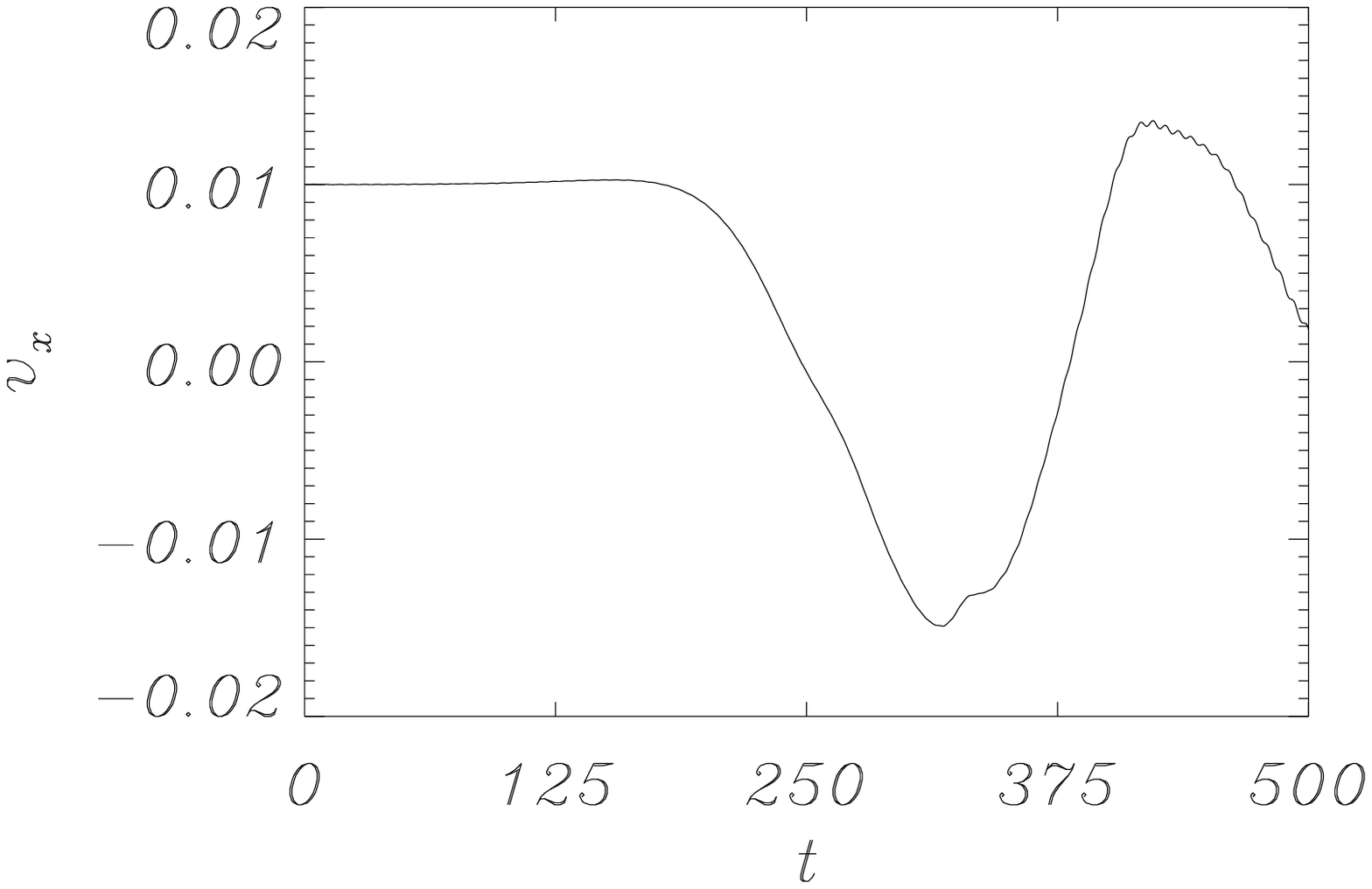,height=5cm,width=5cm}
\psfig{figure=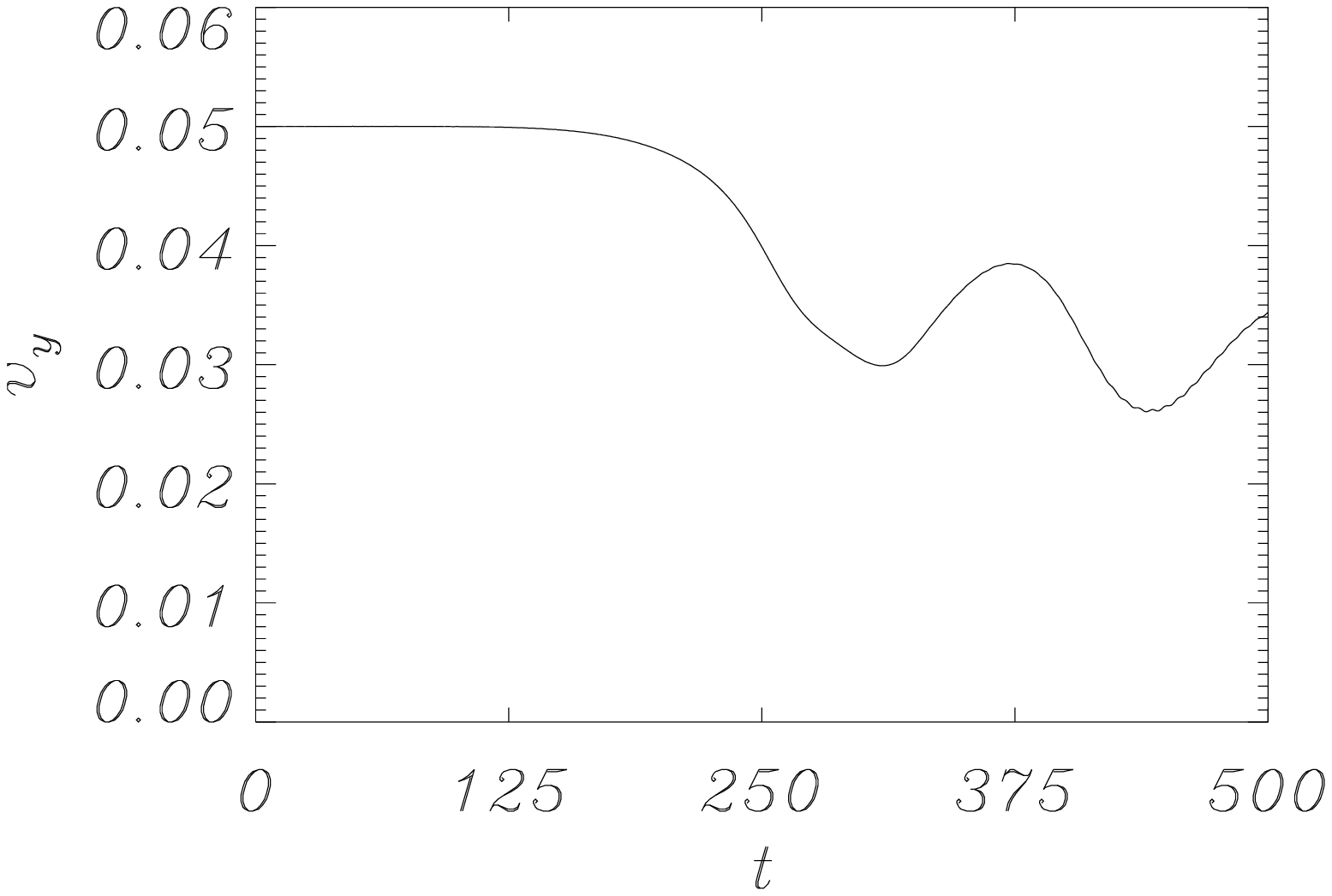,height=5cm,width=5cm}}
\centerline{\psfig{figure=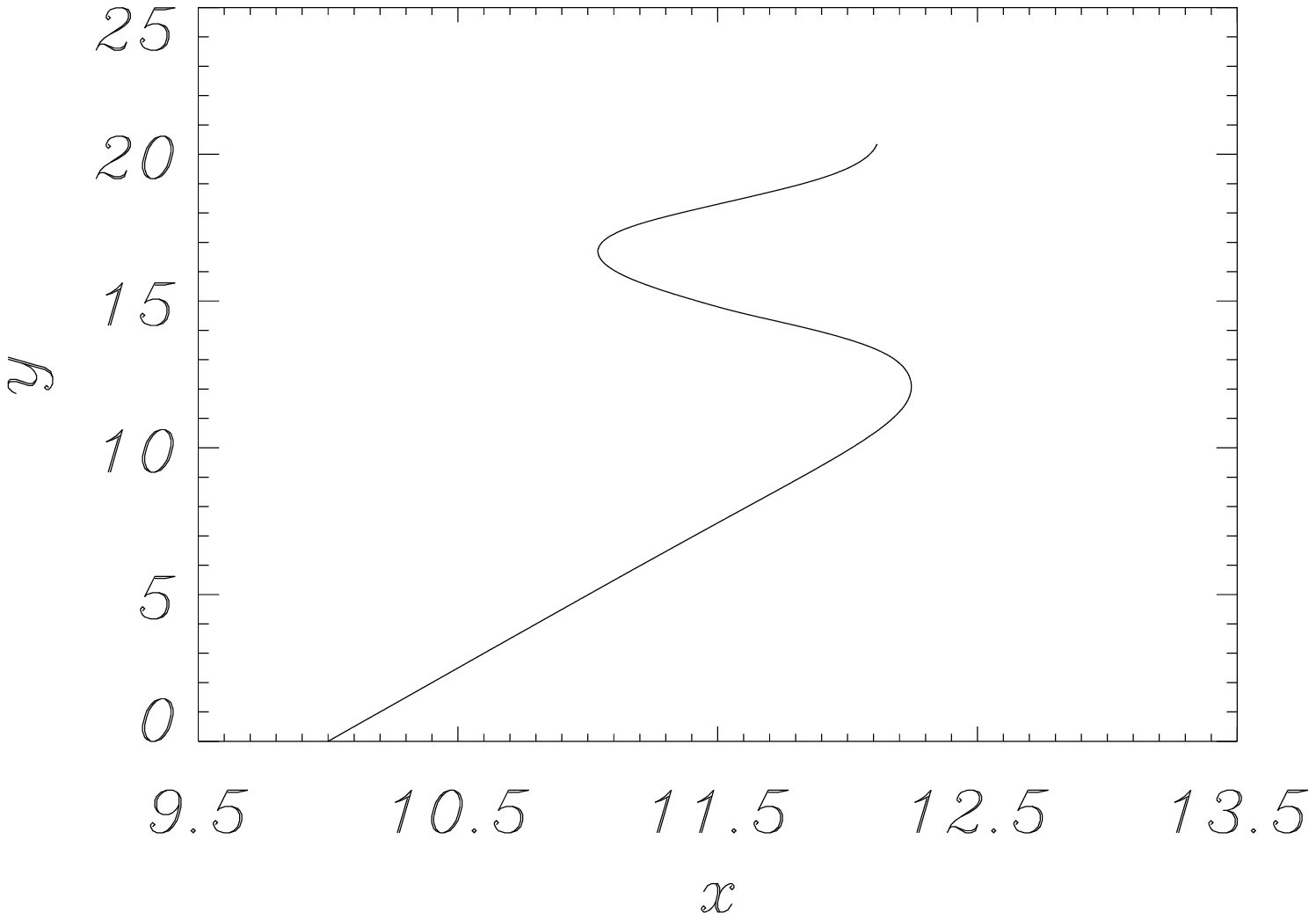,height=5cm,width=5cm}
\psfig{figure=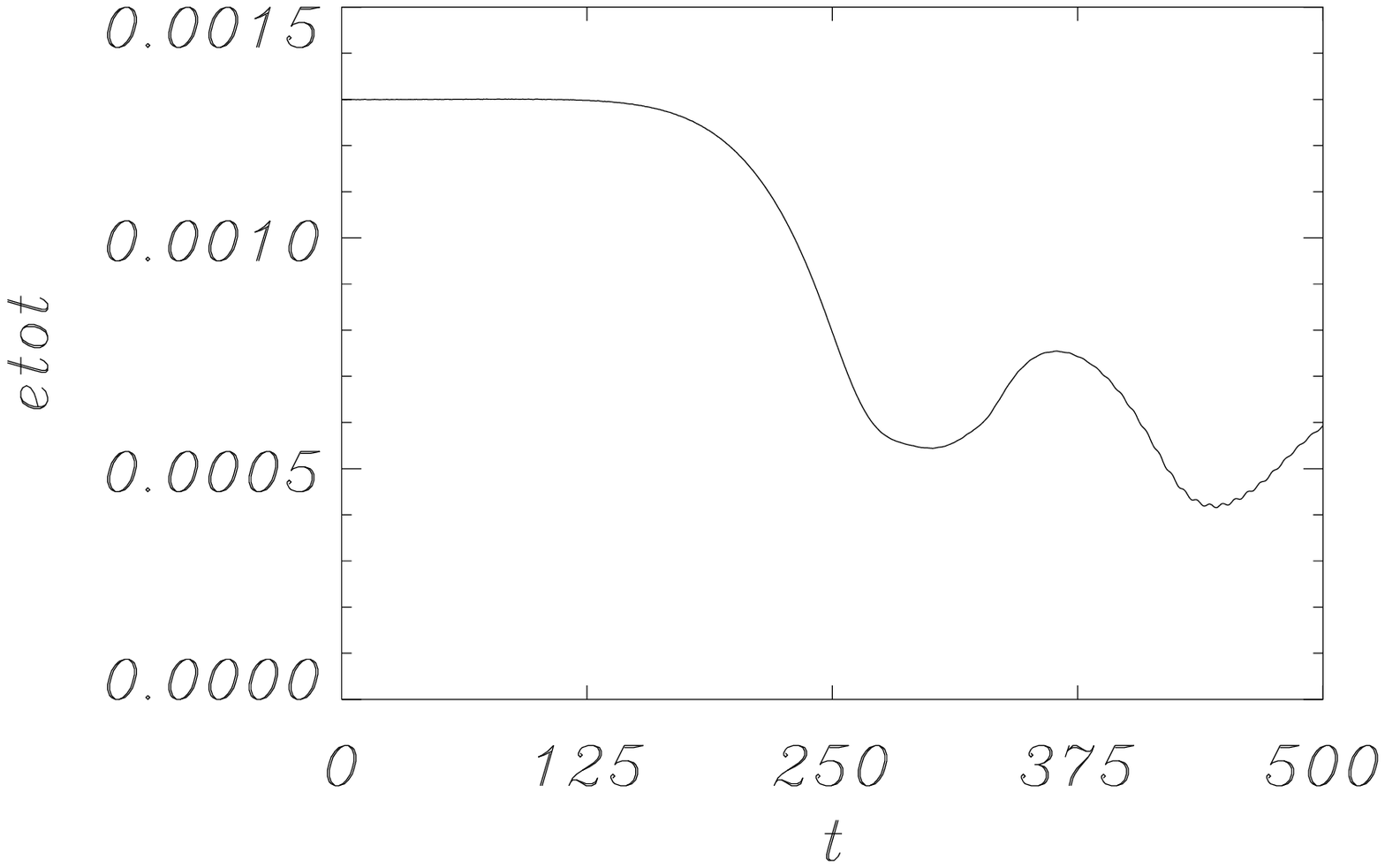,height=5cm,width=5cm}}
\caption[ ]{\small    Example of an electron orbit in the time evolving electromagnetic fields:  particle velocities vs. time (top frames), particle trajectory (bottom left frame), particle kinetic energy vs. time (bottom right frame). }
\label{Fig8}
\end{figure}

For illustration  a representative  orbit is  shown in Fig. \ref{Fig8}. The particle is initialized  at $x = 10$ with  $v_{0x} = 0.01$  and  $v_{0y}= 0.05$  and moves straight until around $t=270$ the  electromagnetic fields  of the instability  have grown significantly.
At this time  this particle  undergoes a net  deceleration  and  its kinetic energy decreases  while $v_{0x}$ and  $v_{0y}$ become oscillatory. The same oscillatory behaviour is displayed by the particle kinetic energy.  The particle trajectory becomes trapped  along $x$ between  two positions, separated by roughly half a wavelength of the dominant $k$ component of $B_z$, where the amplitude of the magnetic field is maximum ($B_z$ is negative near $x = 10.5$ and positive near  $x = 12.5$).

\begin{figure}[!t]
\centerline{\psfig{figure=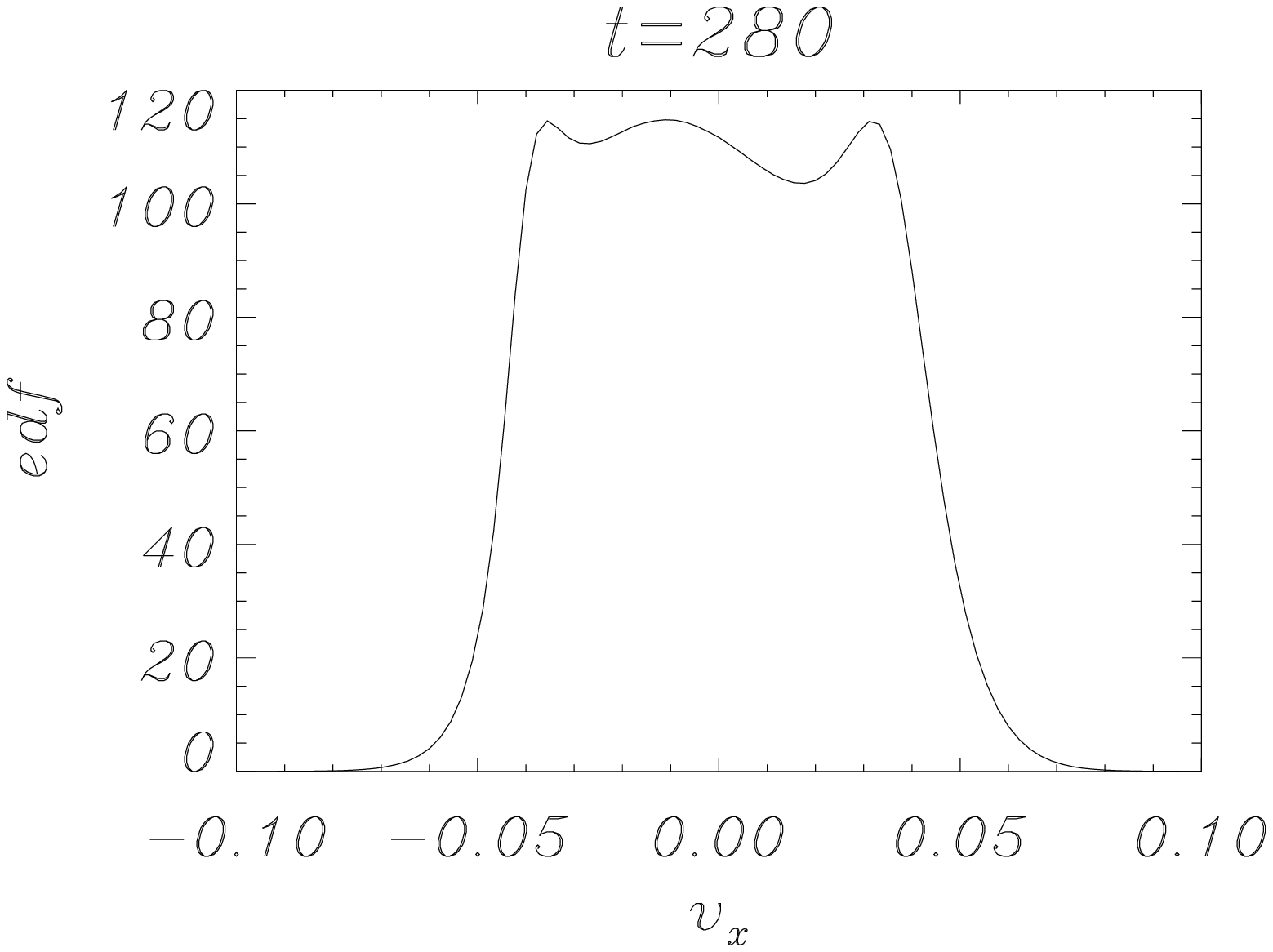,height=5cm,width=5cm}
\psfig{figure=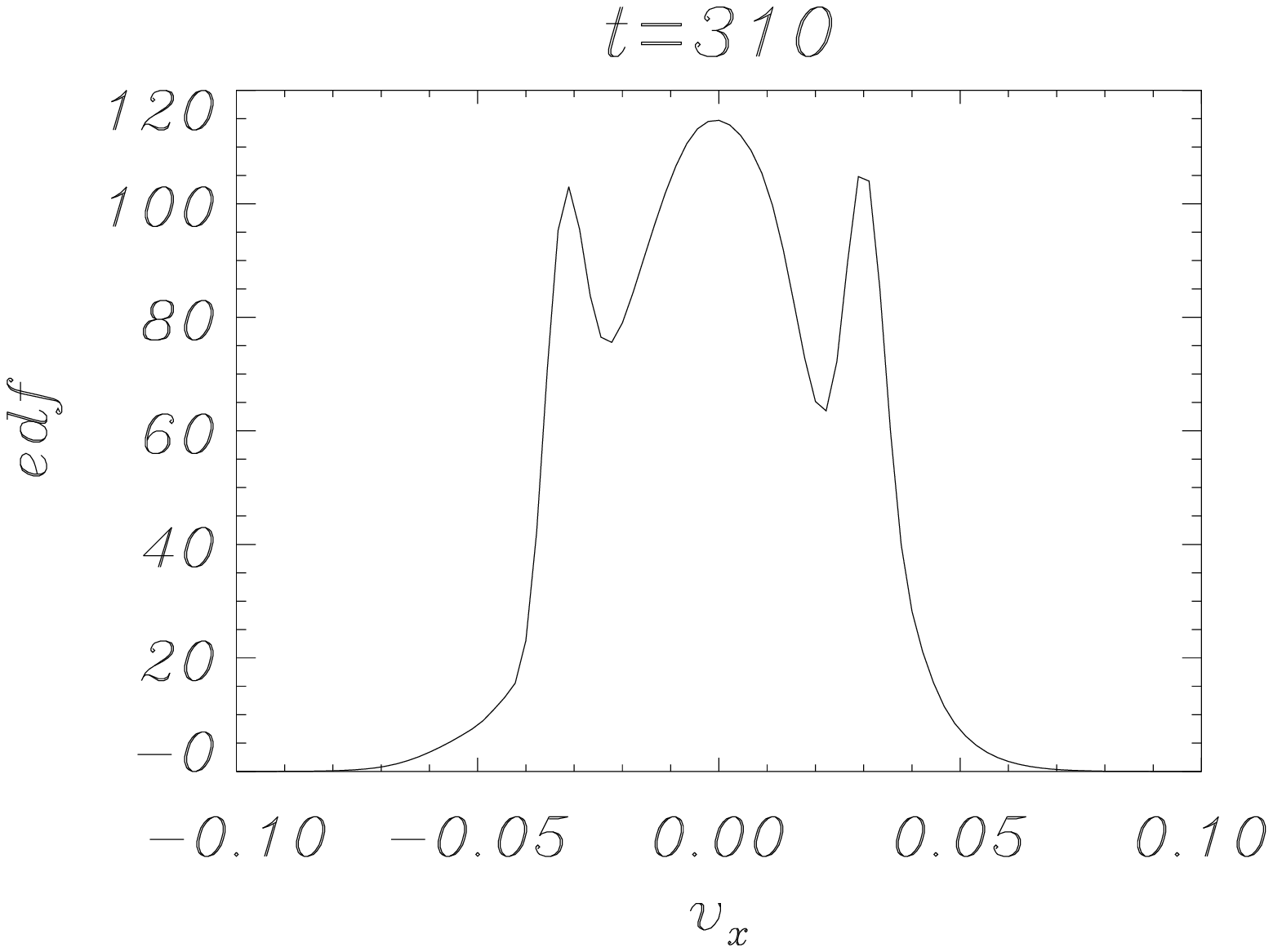,height=5cm,width=5cm}}
\centerline{\psfig{figure=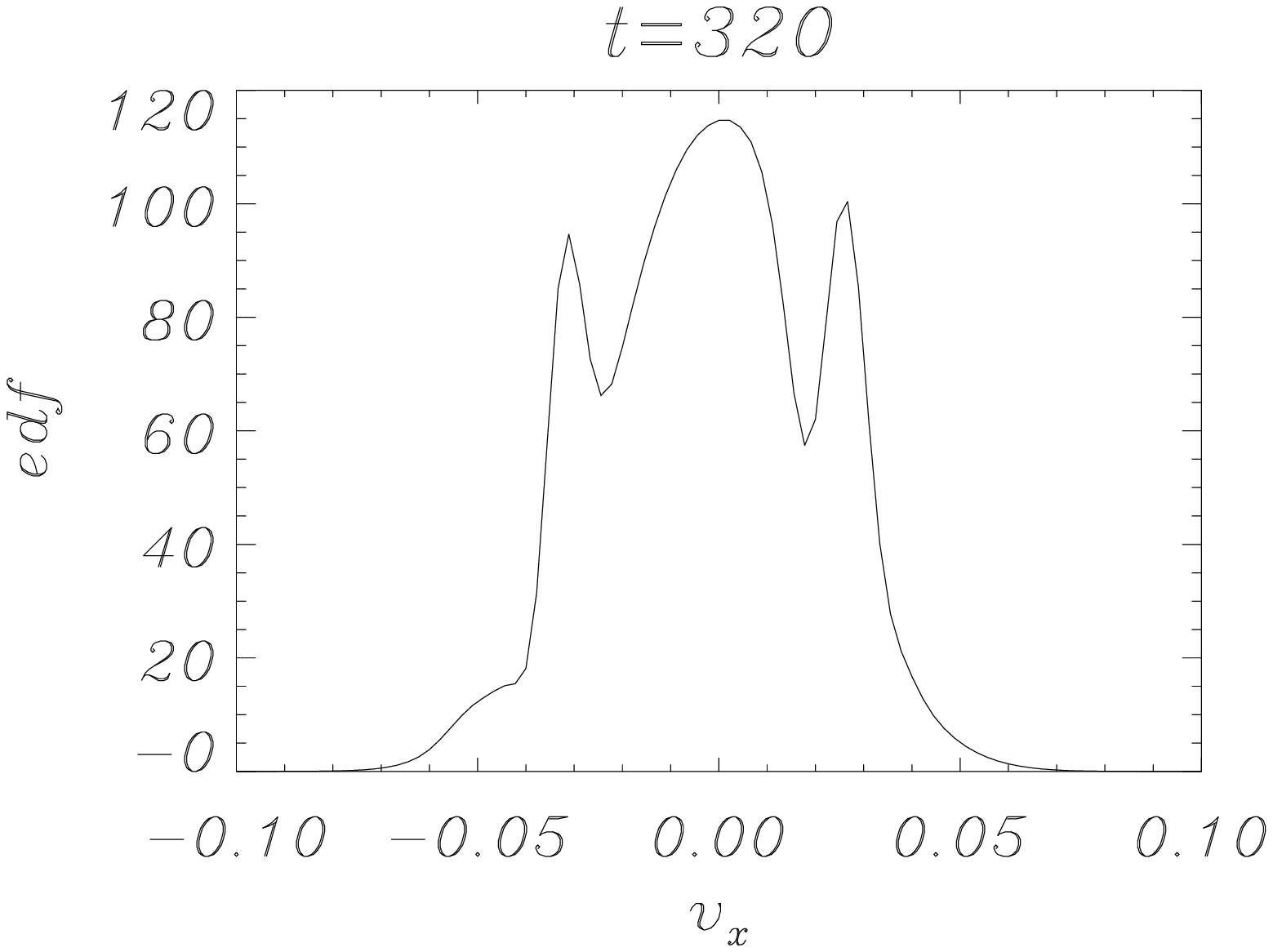,height=5cm,width=5cm}
\psfig{figure=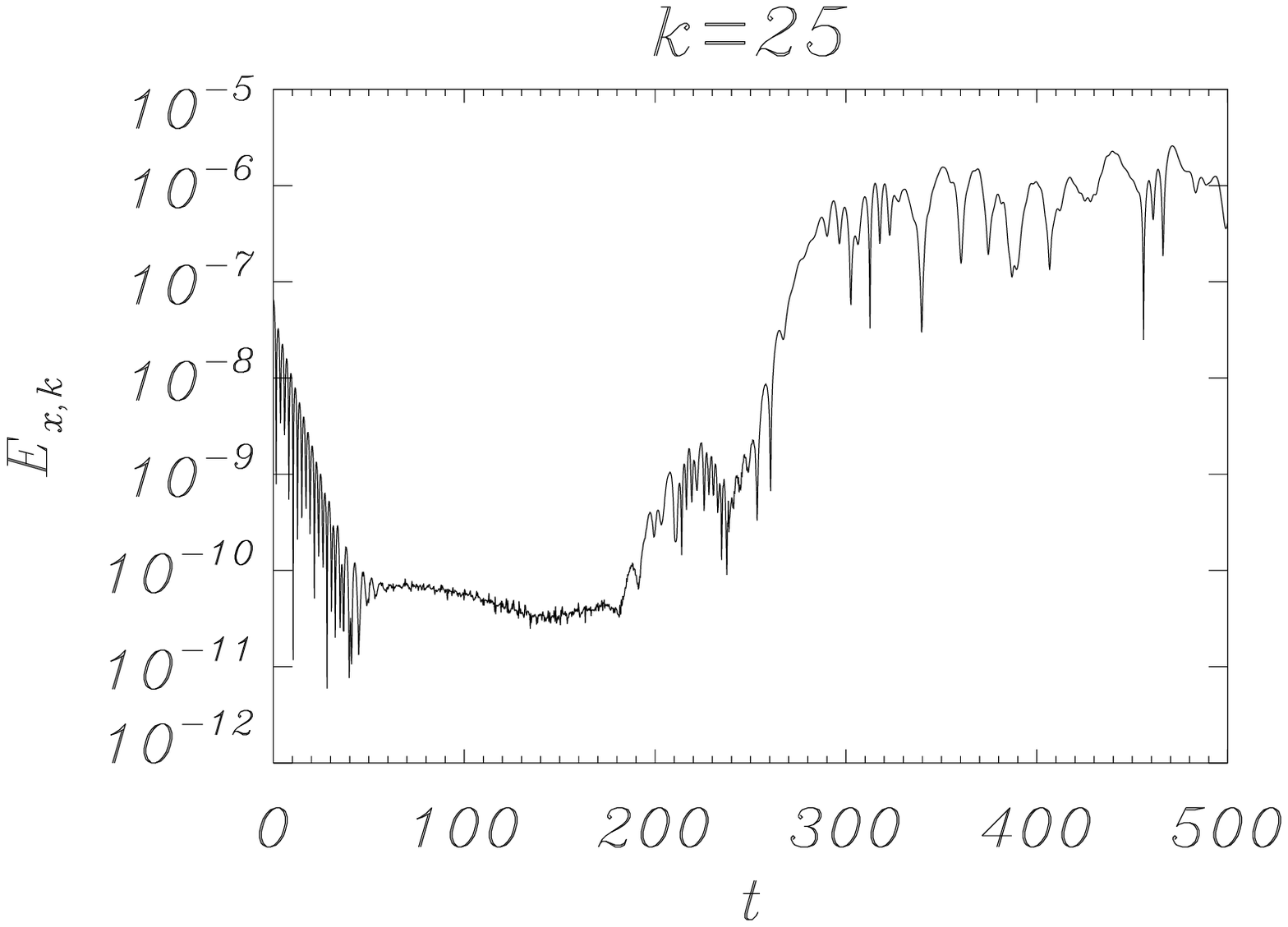,height=5cm,width=5cm}}
\caption[ ]{\small   Electron distribution function at $x=7.8$ and $v_y =0$ vs. $v_x$ at  $t = 280,310,320$ (first three frames).  Time evolution of the longitudinal  electric field component $E_{x, k=25} $ (last frame). Electron distribution function at $t=0$ in the first frame with dashed lines.}
\label{Fig9}
\end{figure}

As shown by the contour plots in  Fig. \ref{Fig5} the electron distribution function becomes ''multi-armed'' as the instability evolves. This implies  that  the distribution function  develops steep positive slopes  for velocities  of the order of  the electron thermal velocity  along $x$ as shown in  Fig. \ref{Fig9}.
Although these positive slopes  evolve in time as the electron distribution function becomes increasingly twisted, they   can
 give rise to a  further excitation of Langmuir waves but   with phase velocities much smaller than those of the Langmuir
 waves driven  by the nonlinear coupling discussed in Sec. \ref{onset}. This new  destabilizing  process is shown in Fig. \ref{Fig9}, last frame,   where the time evolution of the  longitudinal electric field component is shown versus time for $k = 25$  which corresponds to a phase velocity  along $x$ equal to $0.04$.  Contrary to  the components of the longitudinal field  with large phase velocities  shown in Fig. \ref{Fig2}, in the initial phase this high-$k$,  low phase velocity  longitudinal field decays   exponentially to noise level, as consistent with the strong Landau damping due to the initial Maxwellian shape of the electron distribution function.   Later,  around $t= 280$,  the   non monotonic feature  of the electron distribution function shown in the  first frame in  Fig. \ref{Fig9} makes the mode grow  resonantly with a growth rate of the order of $\gamma = 0.15$, i.e. faster than the low-$k$ modes produced by nonlinear density modulations. Eventually  the mode reaches  an amplitude  that is  smaller by approximately an order of magnitude  than that  of  the low $k$ longitudinal modes.

A similar growth  is also exhibited   by longitudinal modes within a  wide range of values of $k$ that extends  towards larger  values of $k$ and thus towards smaller phase velocities that resonate well inside the electron distribution function. The growth of these lower phase velocity modes  starts somewhat earlier than that of the $k= 25$ shown in  Fig. \ref{Fig9} but reaches significantly  lower final  amplitudes.

This destabilization  mechanism shows that  in a  collisionless plasma  dominated by kinetic  effects the nonlinear energy transfer between modes with different wave-number can occur nonlocally in $k$ space, with modes with higher values of $k$ being excited earlier than  modes with comparatively lower values of $k$.

{\bf As mentioned before,  smaller values of the magnetic field $B_z$  are found when smaller  anisotropy ratios are considered. In this case the twisting of the electron distribution function is less pronounced and,  consequently, the excitation of  Langmuir waves  with phase velocities resonating inside  the electron distribution function is considerably weaker.}

\section{Formation of electrostatic coherent structures}

The wide spectrum of longitudinal modes that are excited  by the phase space resonances discussed above leads to a highly structured electron density distribution $n_e$ along $x$,  while the $k$ spectrum magnetic field generated by the Weibel instability is considerably narrower and $B_z$ remains spatially regular even at late times {\bf (see Fig. \ref{Fig7}, left frame).
\begin{figure}[!t]
\centerline{\psfig{figure=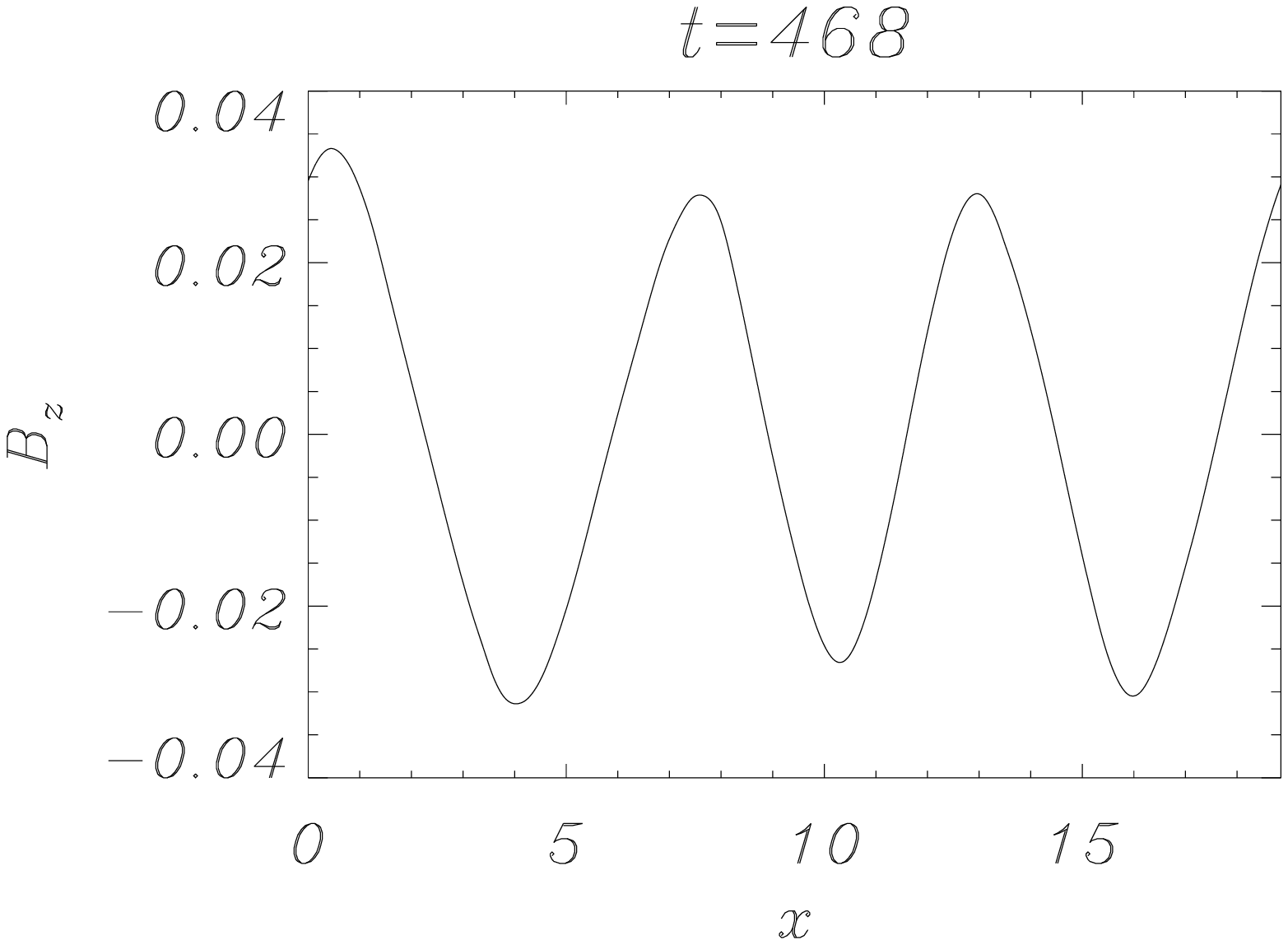,height=5cm,width=7cm}
\psfig{figure=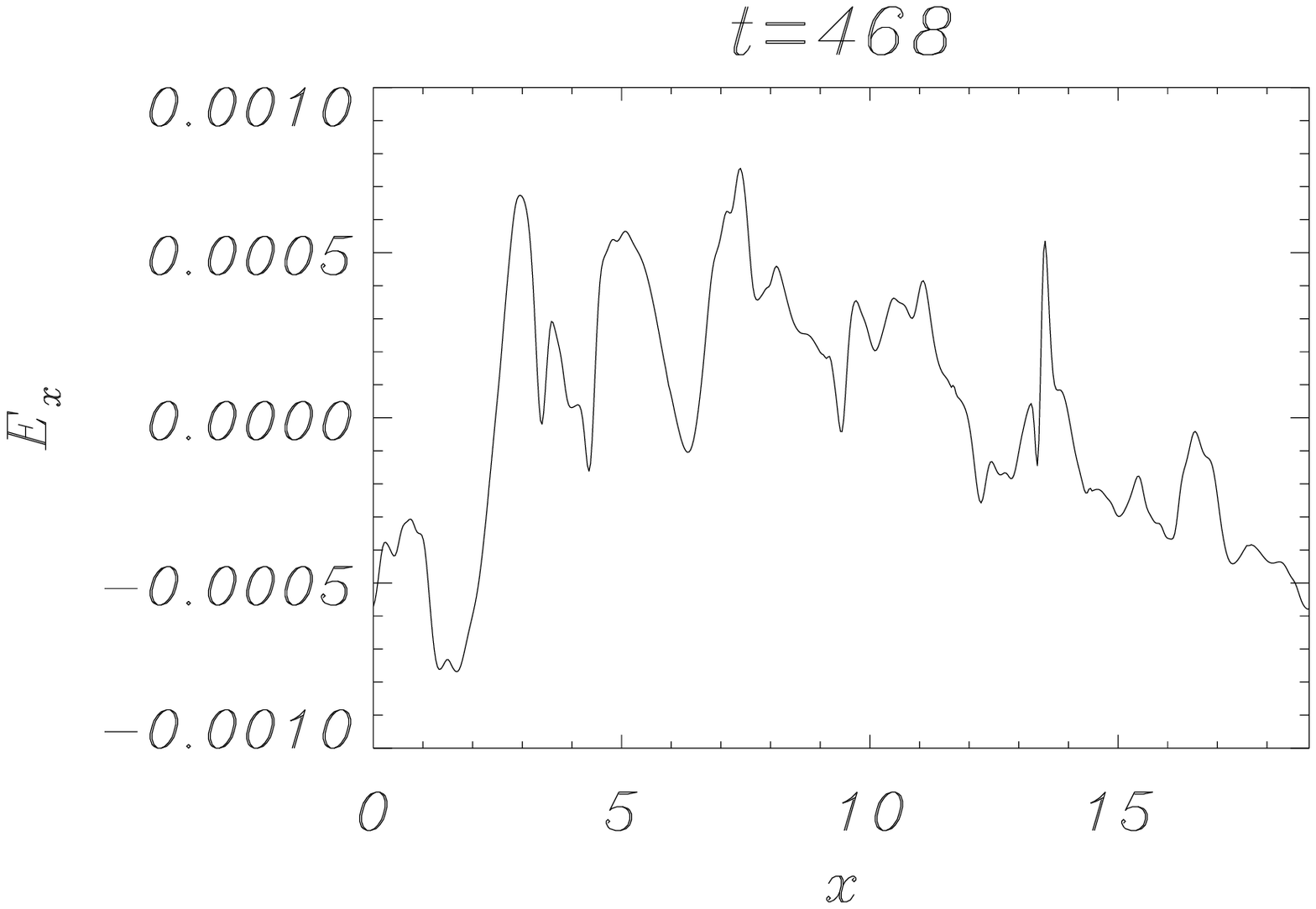,height=5cm,width=7cm}}
\caption[ ]{\small Plot of $B_z$and $E_x$, left and right frame versus $x$ at $t =468$. }
\label{Fig10}
\end{figure}
This is also illustrated in  Fig. \ref{Fig10} where we show the spatial profile of $B_z$ and $E_x$} versus $x$ at $t = 468$. It is seen that the electric field, besides exhibiting a spatial modulation at roughly  half the dominant magnetic field wavelength, has developed {\bf several narrow spikes, in particular a very narrow one at $x\sim 13.6$ (corresponding to a sharp minimum  in the  the electron density). In the one-dimensional configuration considered here, these  spikes correspond to "multi layer" electrostatic structures accompanied by a drop in
the potential. A detailed analysis shows that these structures are not a true equilibrium between the electrostatic field and the pressure gradient, as is generally  found for double layer coherent structures associated with phase space holes Ref.  \cite{cal05}. However, since these structures survive on time scales of the order of several ion periods, we still define them as coherent electrostatic structures.}
The time evolution of  the  longitudinal electric field is shown in the first four frames in Fig. \ref{Fig11} vs. $x$ and in the  last frames in Fig. \ref{Fig11} where  $E_x$ at $x=13.6$ is plotted vs.  $t$. A clear view of the formation of the electrostatic structures is given in Fig. \ref{Fig12} where we show the electrostatic field in the ($x,t$) plane. We see that after forming large scale structures on the magnetic (one-half) scale around $t=250$, slowly propagating small scales electrostatic multipolar structures are formed on the Debye length scale.
As they  propagate, these structures interact with each other leading to coalescence or to the formation of new structures with similar characteristics.\,\, 
{\bf  We observe that these structures do not form in the numerical simulations with  smaller 
anisotropy ratios ($T_y/T_x = 4$).  }
\begin{figure}[!t]
\centerline{\psfig{figure=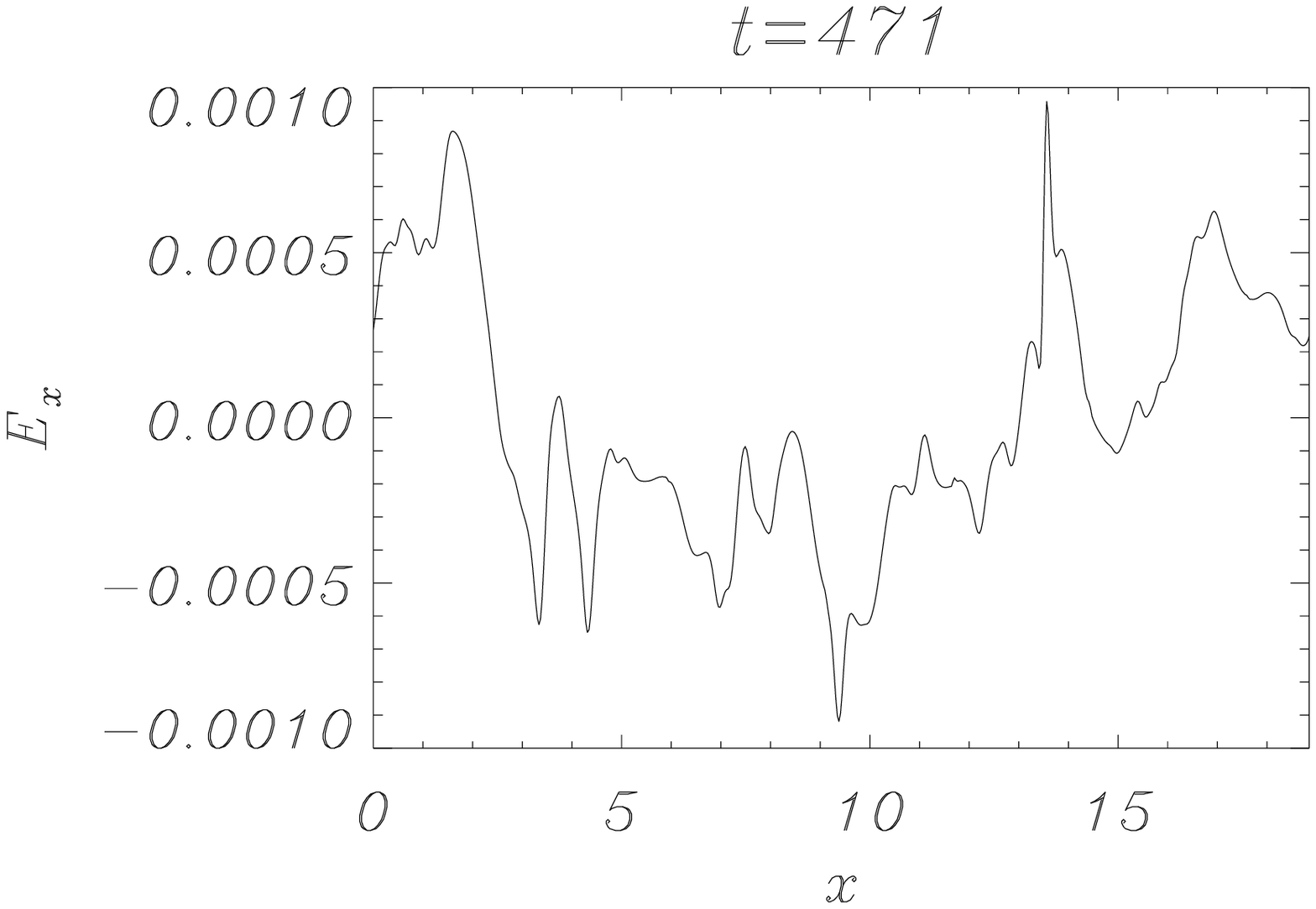,height=5cm,width=7cm}
\psfig{figure=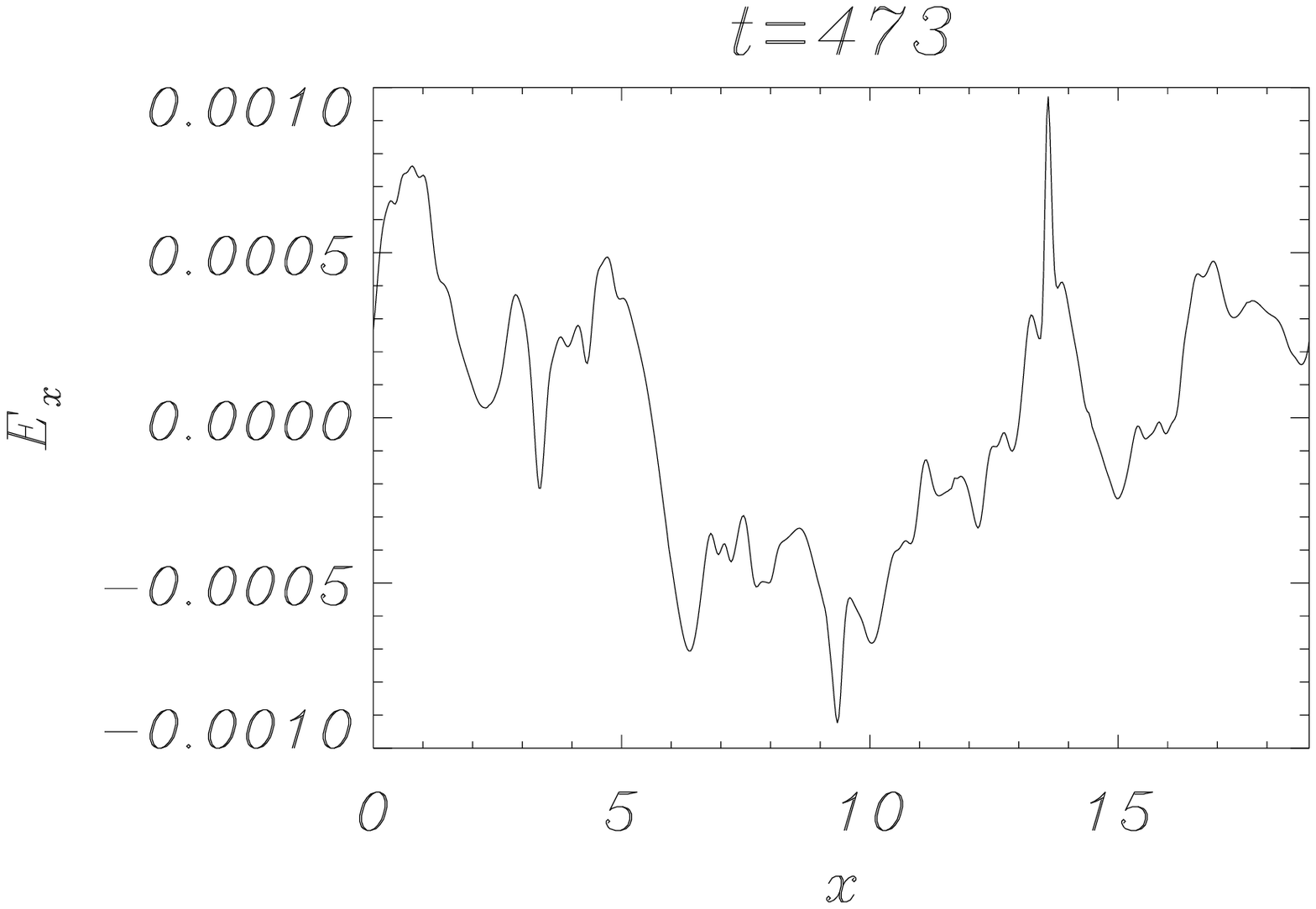,height=5cm,width=7cm}}
\centerline{\psfig{figure=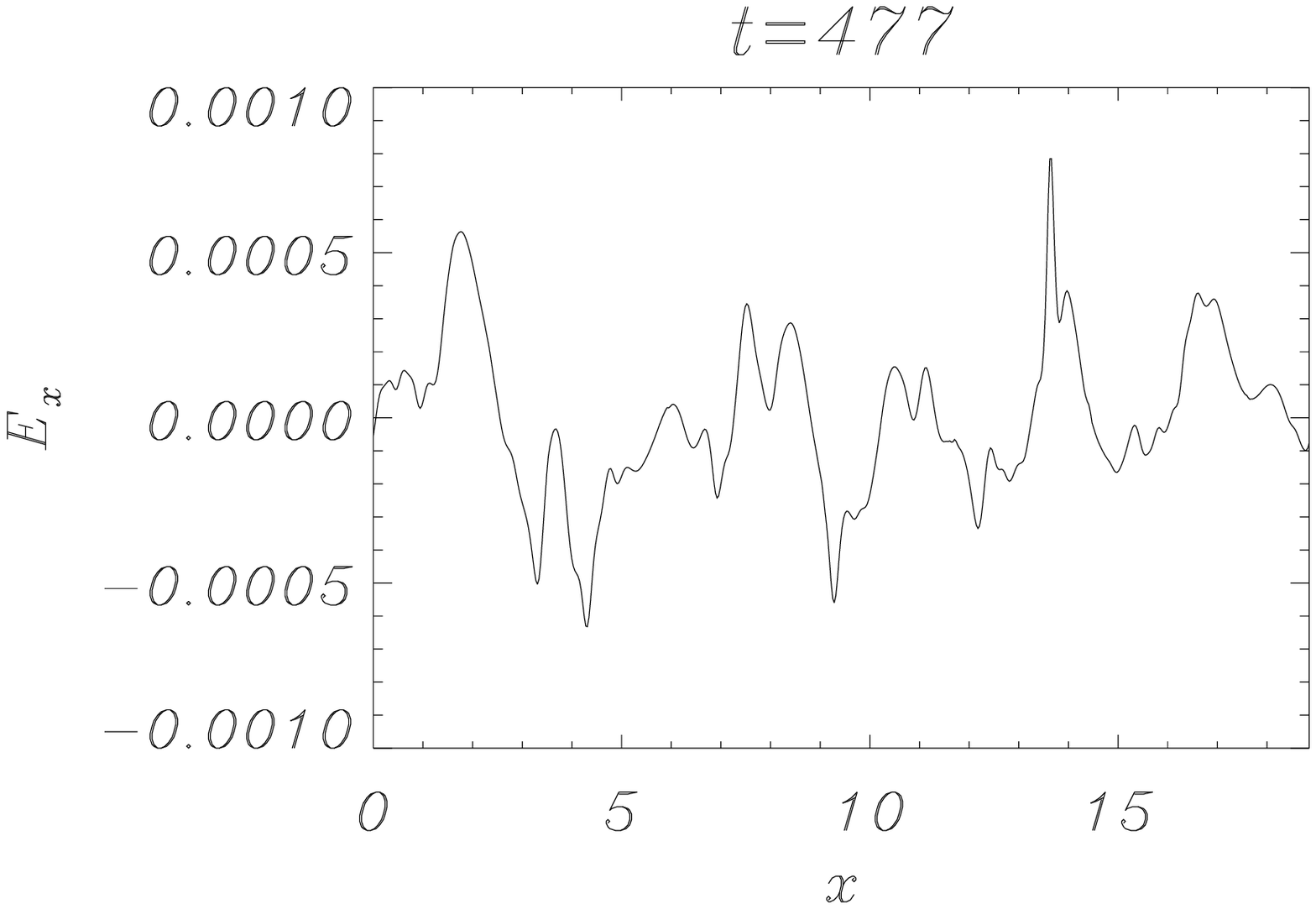,height=5cm,width=7cm}
\psfig{figure=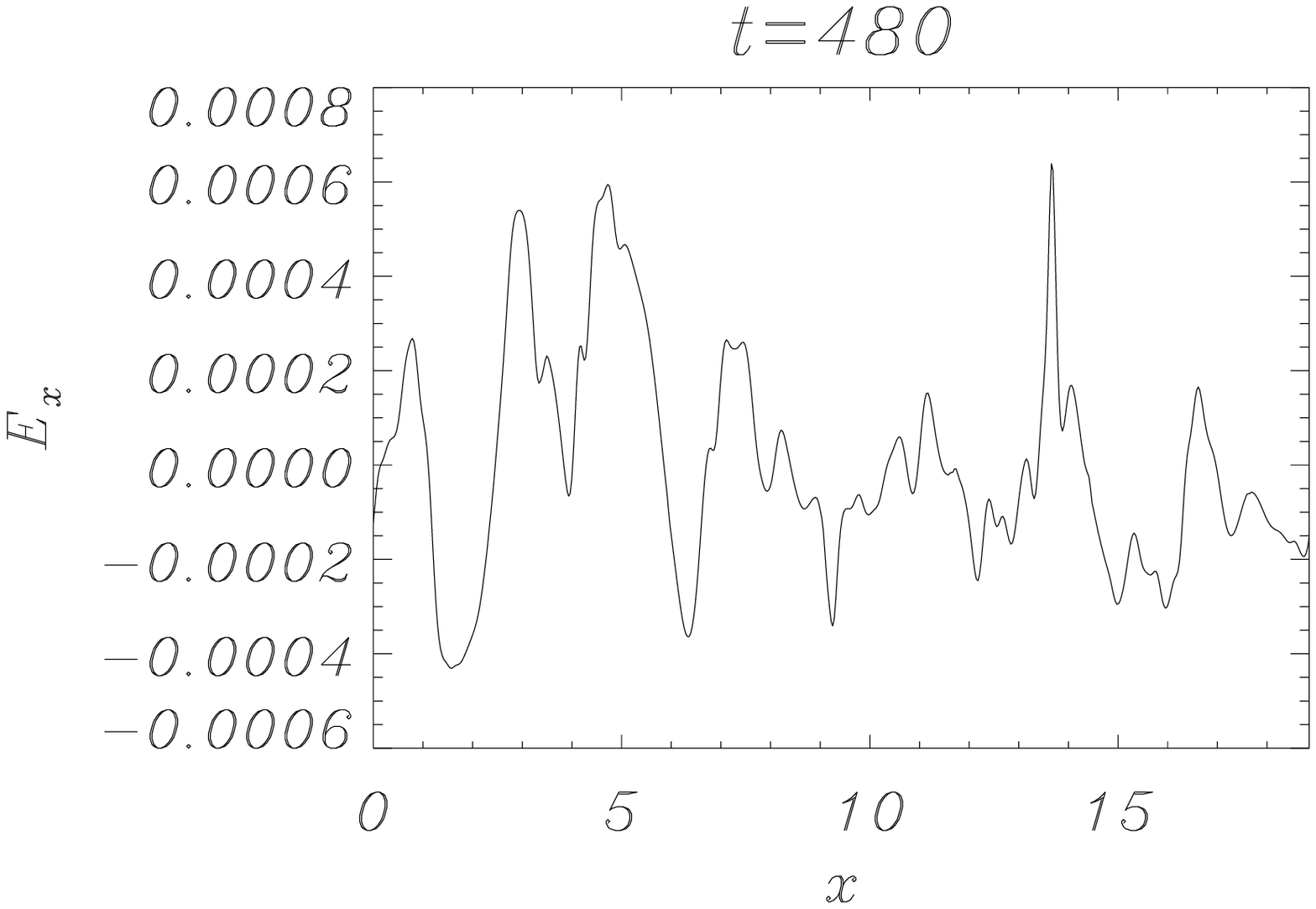,height=5cm,width=7cm}}
\centerline{\psfig{figure=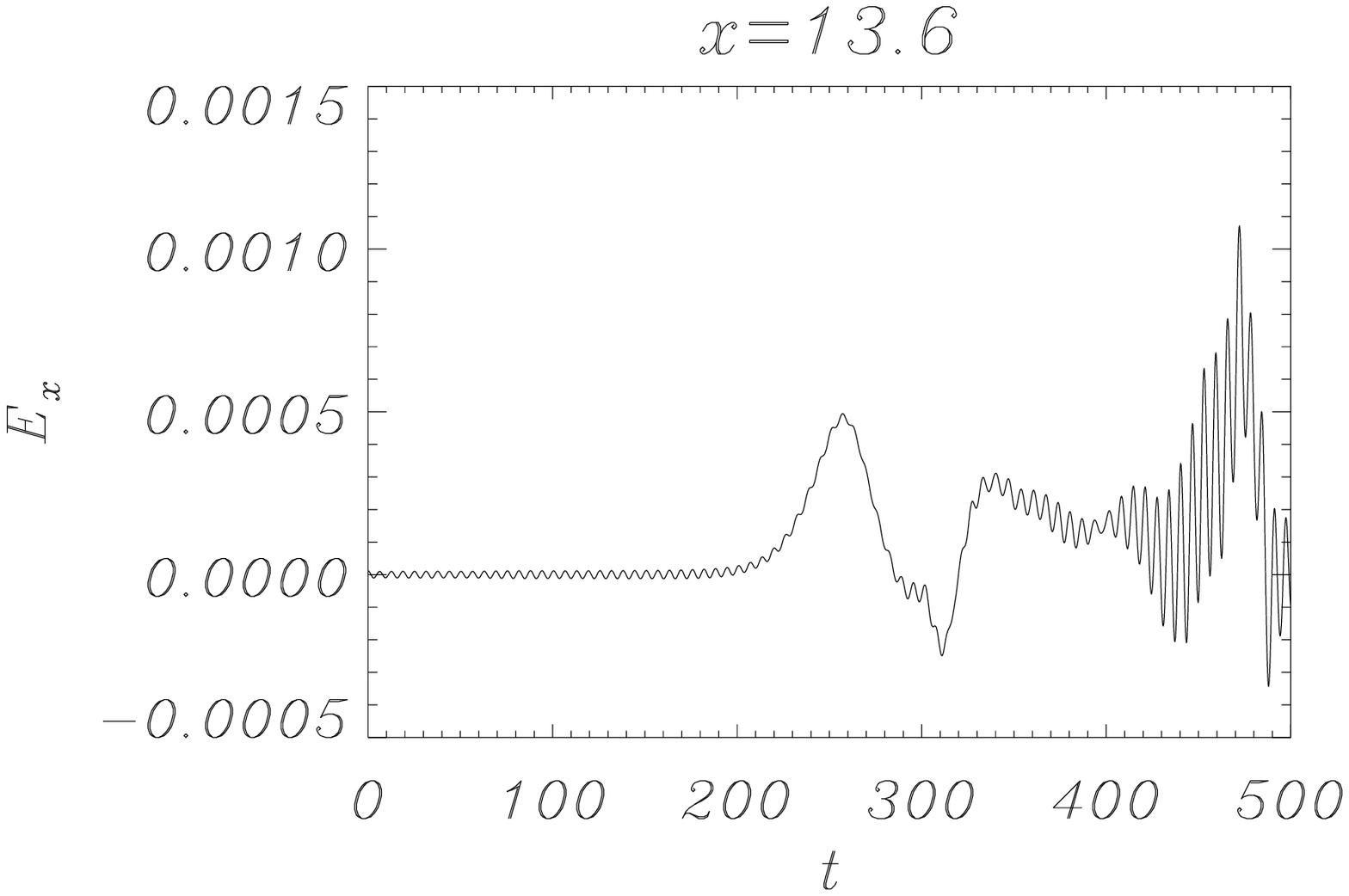,height=5cm,width=7cm}
\psfig{figure=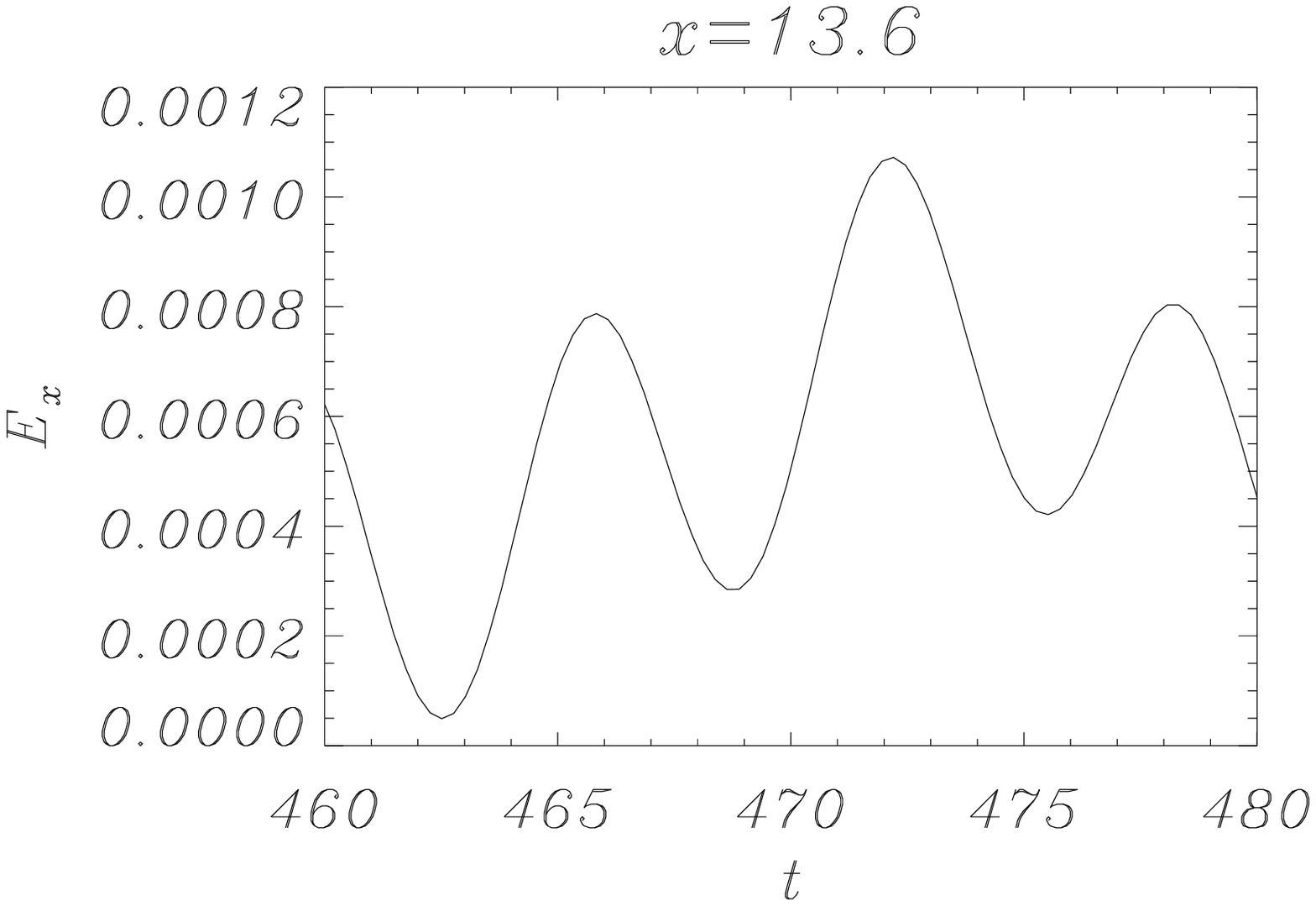,height=5cm,width=7cm}}
\caption[ ]{\small {\bf Top four frames: Plot of $E_x$ versus $x$ at $t =471,473,477,480$. Last two frames: Plot of $E_x$ versus $t$ at $x=13.6$} } 
\label{Fig11}
\end{figure}

\begin{figure}[!t]
\centerline{\psfig{figure=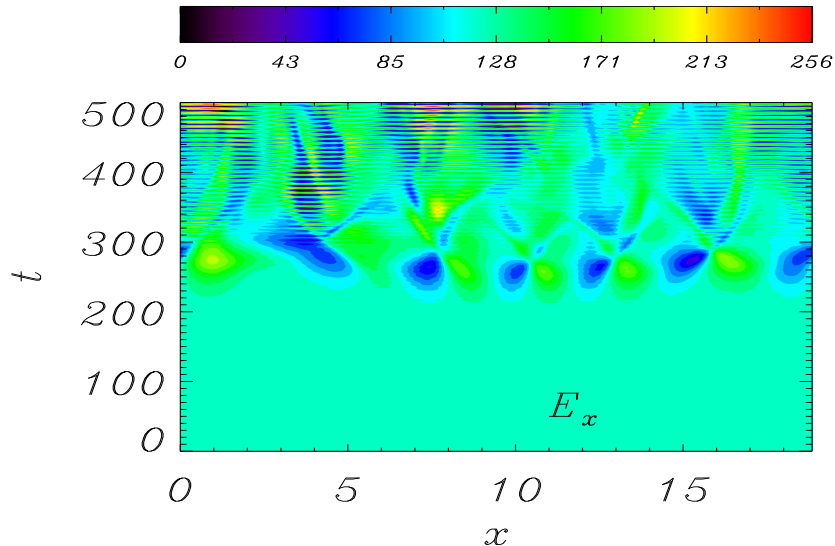,height=8cm,width=14cm}}
\caption[ ]{\small Contour plots in the ($x-t$) plane of the {\bf  electrostatic }  field $E_x$.}
\label{Fig12}
\end{figure}

Electrostatic features of this type are encountered under space conditions and were first observed almost thirty years ago in auroral regions in the form of dipolar electric structures as well as double layers \cite{Temerin}. These "large amplitude solitary structures" are in general observed with the electric field aligned with the mean magnetic field direction \cite{ergun}. In the solar wind observations of dipolar structure have been reported starting from ten years ago in Ref. \cite{mangeney}. From a theoretical point of view, there is no consensus about the origin of such structures. Accelerated particles, or beams, generated far away by some energetic events such  as, for example, magnetic reconnection, are probably the best candidate for  the energetic driver of the mechanism at the basis of the formation of multipolar structures \cite{matsumoto,omura,briandjgr}. In this context, it is interesting to note that plasma temperature anisotropies, a typical feature expected in many regions in solar wind plasmas, may represent a possible source for their formation through a multi stage process that involves a complex evolution of the electron distribution function under the influence of the electromagnetic fields generated by the Weibel instability. A study of plasma electron temperature anisotropy in the presence of a weak, mean magnetic field (see \cite{gary1} for the linear analysis and \cite{gary2} for simulations) is in progress.

\section{Conclusions}

Pressure anisotropy is a common feature of laboratories and of space plasma and provides the free energy source for the development  of the Weibel instability and the generation of a quasi static magnetic field.  In this paper we have investigated the nonlinear evolution of the Weibel instability arising from the anisotropy of the electron distribution function in a collisionless plasma.   We have adopted a  two-dimensional velocity space  Vlasov code in a simplified spatially one-dimensional configuration.

We have  found that the  instability causes  a violent deformation of the electron distribution function in phase space leading to the generation of short wavelength Langmuir modes. This  mechanism coexists  with the ''fluid-like'' generation of longer wavelength Langmuir modes  due to the nonlinear modulation of the electron density. It corresponds to a non-local  energy transport  between modes in wave-number space.  Eventually  these short wavelength Langmuir modes lead to the formation  of highly  localized electrostatic structures.  These structures correspond to  potential jumps and are of interest for space observations.

\end{document}